\documentclass[aps,pra,twocolumn]{revtex4-1}

\usepackage[T1]{fontenc}
\usepackage{amsmath,amsfonts,amssymb,amsthm,bbm,bm, braket}
\usepackage{wasysym}
\usepackage{comment}
\usepackage{enumerate}
\usepackage{graphicx}
\usepackage{mathtools}
\usepackage{ifthen}
\usepackage{tensor}
\usepackage{tikz}
\usepackage{tikz-network}
\usetikzlibrary{patterns,decorations.pathreplacing}
\theoremstyle{break}        

\theoremstyle{break}        

\usepackage{color}
\definecolor{myred}{RGB}{232,102,102}
\definecolor{myblue}{RGB}{187,187,255}
\definecolor{mygreen}{RGB}{34,139,34}
\definecolor{myorange}{RGB}{255,165,0}
\definecolor{OliveGreen}{RGB}{85,107,47}
\definecolor{NavyBlue}{RGB}{0,0,128}
\usepackage{tensor}
\newcommand{\be}{\begin{equation}}
\newcommand{\ee}{\end{equation}}
\newcommand{\ba}{\begin{aligned}}
\newcommand{\ea}{\end{aligned}}
\newcommand{\1}{\mathbbm{1}}

\theoremstyle{plain}
\newtheorem{property}{Property}
\theoremstyle{plain}

\usepackage[colorlinks,bookmarks=false,citecolor=NavyBlue,linkcolor=OliveGreen,urlcolor=blue]{hyperref}

\begin{document}

\title{Exact Correlation Functions for Dual-Unitary Lattice Models in 1+1 Dimensions}

\author{Bruno Bertini, Pavel Kos, and Toma\v z Prosen}

\affiliation{Department of Physics, Faculty of Mathematics and Physics, University of Ljubljana, Jadranska 19, SI-1000 Ljubljana, Slovenia}

\date{\today}

\begin{abstract}
We consider a class of quantum lattice models in $1+1$ dimensions represented as local quantum circuits that enjoy a particular `dual-unitarity' property. In essence, this property ensures that both the evolution `in time' and that `in space' are given in terms of unitary transfer matrices. We show that for this class of circuits, generically non-integrable, one can compute explicitly all dynamical correlations of local observables. Our result is exact, non-pertubative, and holds for any dimension $d$ of the local Hilbert space. In the minimal case of qubits ($d = 2$) we also present a classification of all dual-unitary circuits which allows us to single out a number of distinct classes for the behaviour of the dynamical correlations.
We find `non-interacting' classes, where all correlations are preserved, the ergodic and mixing one, where all correlations decay, and, interestingly, also classes that are are both interacting and non-ergodic.

\end{abstract}
\maketitle

Spatiotemporal correlation functions of local observables provide the most fundamental and useful physical description of locally interacting classical or quantum many-body systems~\cite{altland,sethna}. They characterise ergodic properties~\cite{arnold}, as well as basic transport coefficients of many body systems, such as conductivities, Drude weights, and kinematic viscosities~\cite{mahanbook}. Moreover, correlation functions are directly measurable quantities, e.g., via X-ray diffraction or neutron scattering in solid state materials~\cite{mahanbook}, or via quantum optical detection techniques in cold atomic gases~\cite{blockreview}.

In spite of their extraordinary relevance, an exact non-perturbative calculation of correlation functions is generically feasible only in free (Gaussian) theories, or for certain types of completely integrable models \cite{korepin}. To date, no specific example of strongly coupled, non-integrable, local quantum many-body system with accessible correlation functions is known. 

The situation is somewhat different for classical dynamical systems with few degrees of freedom, where some exactly tractable examples have been found. For instance, Arnold's cat map, Baker's map, and dispersing billiards~\cite{arnold,cornfeld,gaspard,ott}. In classical dynamical systems the decay of correlation functions leads to a rigorous criterion of ergodicity and dynamical mixing \cite{cornfeld}. Moreover, the presence of tractable systems allows for rigorous results on quantum eigenstate ergodicity~\cite{Zelditch,ColinDeVerdiere,DeBievre}. On the contrary, ergodic theory of quantum many-body systems is currently at its early infancy. In particular, in locally interacting systems it is exceptionally challenging to separate relaxation (mixing, or `scrambling') mechanisms from the mere decaying time correlations of local operators. The former requires dynamical complexity and thus absence of integrability~\cite{TP_PRE99,KI_JPA,ehud}, while the latter occurs even in quasi-free systems~\cite{lenci,graffi,KI_JPA}. 

Here we make the first step towards a rigorous ergodic theory for quantum many-body systems by identifying a class of quantum lattice models in one spatial dimension with explicitly accessible spatiotemporal correlation functions for arbitrary pairs of ultra-local observables. These models are generically non-integrable and can be formulated in terms of local quantum circuits with local gates exhibiting a particular `dual-unitarity' property.  Specifically, they are unitary and remain unitary under a reshuffling of their indices, ensuring that the full quantum circuit defines unitary evolutions in both time and space directions. This class of models includes both the non-integrable self-dual kicked Ising model (SDKI), where the aforementioned feature recently enabled us to find exact results on spectral correlations and entanglement entropy dynamics~\cite{letter,entropy}, and some integrable Floquet systems~\cite{PoGri17, Lenart1, Lenart2}.

More specifically, we consider quantum systems defined on a periodic chain of $2L$ sites, where each site is equipped with a $d$-dimensional local Hilbert space 
$\mathcal H_1= {\mathbb C}^d$ with a basis ${\{\ket{i};\,i=1,2\ldots d\}}$; the Hilbert space of the system is then ${\cal H}_{2L}= \mathcal H_1^{\otimes 2L}$.
The time evolution is discrete and local. In particular, each time step is divided in two halves. In the first half the system is evolved by the transfer matrix     
$
\mathbb U^{\rm e}= U^{\otimes L}
$,
where ${U\in {\rm End}(\mathcal H_1 \otimes \mathcal H_1)}$ is the `local gate' and encodes the physical properties of the system. In the second half, instead, the system is evolved by 
$
\mathbb U^{\rm o}=\mathbb T^{\phantom{\dag}}_{2L} \mathbb U^{\rm e} \mathbb T^\dag_{2L}
$, where ${\mathbb T_\ell \ket{i_1}\otimes\ket{i_2}\cdots\ket{i_{\ell}}\equiv \ket{i_2}\otimes \ket{i_3} \cdots \ket{i_\ell}\otimes \ket{i_1}}$ is a $\ell-$periodic translation by one site.
This means that the transfer matrix for an entire time step is given by  
\be
\mathbb U=\mathbb U^{\rm o}\mathbb U^{\rm e} = \mathbb T^{\phantom{\dag}}_{2L} U^{\otimes L} \mathbb T^\dag_{2L} U^{\otimes L}\,.
\label{eq:timetransfermatrix}
\ee
Note that, from the definition \eqref{eq:timetransfermatrix} it immediately follows that $\mathbb U$ is invariant under two-site shifts $\mathbb U \mathbb T^2_{2L}= \mathbb T^2_{2L} \mathbb U$.

Before continuing, we note that the systems under exam admit a convenient diagrammatic representation. One depicts operators as boxes with a number of incoming and outgoing legs corresponding to the number of local sites they act on. Each leg (or wire) carries a Hilbert space $\mathcal H_1$. For instance, operators acting on a single site are represented as a line with a bullet $\bullet$, while the local gate and its Hermitian conjugate are represented as  
\begin{equation}
\label{eq:graphical}
\begin{tikzpicture}[baseline=(current  bounding  box.center), scale=0.8]
\draw[thick] (-4.25,0.5) -- (-3.25,-0.5);
\draw[thick] (-4.25,-0.5) -- (-3.25,0.5);
\draw[thick, fill=myred, rounded corners=2pt] (-4,0.25) rectangle (-3.5,-0.25);
\Text[x=-4.75,y=0.05]{$U=$}
\Text[x=-3.,y=-0.075]{,}
\end{tikzpicture}
\qquad 
\qquad
\begin{tikzpicture}[baseline=(current  bounding  box.center), scale=0.8]
\draw[thick] (-1.25,0.5) -- (-.25,-0.5);
\draw[thick] (-1.25,-0.5) -- (-.25,0.5);
\draw[thick, fill=myblue, rounded corners=2pt] (-1,0.25) rectangle (-0.5,-0.25);
\Text[x=-1.75,y=0.075]{$U^\dag=$}
\Text[x=0,y=-0.075]{.}
\end{tikzpicture}
\end{equation}
We stress that, even if we use a symmetric symbol for $U$, we assume no symmetry under reflection (left to right flip) and time reversal (up to down flip).

Diagrammatic representation allows to depict the trace of the propagator for $t$ steps as a partition function of a certain vertex model
\be
\!\!\!\!\!{\rm tr}\left[\mathbb U^t\right] \!=\!
\begin{tikzpicture}[baseline=(current  bounding  box.center), scale=0.45]
\foreach \i in {0,1,2,3}
{
\Text[x=1.25,y=+2*\i]{\scriptsize$\i$}
}
\foreach \i in {1,3,5}
{
\Text[x=1.25,y=+\i]{\small$\frac{\i}{2}$}
}
\foreach \i in {0,...,2}
{
\Text[x=-6.5+2*\i+2,y=-1]{\scriptsize${\i}$}
}
\foreach \i in {-2,-1}
{
\Text[x=-6.8+2*\i+2,y=-1]{\scriptsize${\i}$}
}
\foreach \i in {1,3,5}
{
\Text[x=-6.5+\i+2,y=-1]{\small$\frac{\i}{2}$}
}
\foreach \i in {3,1}
{
\Text[x=-6.8-\i+2,y=-1]{\small$\,-\frac{\i}{2}$}
}
\foreach \i in {3,...,13}
{
\draw[thick, gray, dashed] (-12.5+\i,-0.1) -- (-12.5+\i,6.3);
}
\foreach \i in {-1,...,5}
{
\draw[thick, gray, dashed] (-9.75,5-\i) -- (.75,5-\i);
}
\foreach \i in {0,...,4}
{
\draw[thick] (-.5-2*\i,1) -- (0.5-2*\i,0);
\draw[thick] (-0.5-2*\i,0) -- (0.5-2*\i,1);
\draw[thick, fill=myred, rounded corners=2pt] (-0.25-2*\i,0.25) rectangle (.25-2*\i,0.75);}
\foreach \i in {1,...,5}
{
\draw[thick] (.5-2*\i,6) -- (1-2*\i,5.5);
\draw[thick] (1.5-2*\i,6) -- (1-2*\i,5.5);
}
\foreach \jj[evaluate=\jj as \j using -2*(ceil(\jj/2)-\jj/2)] in {0,...,3}
\foreach \i in {1,...,5}
{
\draw[thick] (.5-2*\i-1*\j,2+1*\jj) -- (1-2*\i-1*\j,1.5+\jj);
\draw[thick] (1-2*\i-1*\j,1.5+1*\jj) -- (1.5-2*\i-1*\j,2+\jj);
}
\foreach \jj[evaluate=\jj as \j using -2*(ceil(\jj/2)-\jj/2)] in {0,...,4}
\foreach \i in {1,...,5}
{
\draw[thick] (.5-2*\i-1*\j,1+1*\jj) -- (1-2*\i-1*\j,1.5+\jj);
\draw[thick] (1-2*\i-1*\j,1.5+1*\jj) -- (1.5-2*\i-1*\j,1+\jj);
\draw[thick, fill=myred, rounded corners=2pt] (0.75-2*\i-1*\j,1.75+\jj) rectangle (1.25-2*\i-1*\j,1.25+\jj);
}
\Text[x=2,y=5]{$t$}
\Text[x=-5.5,y=-2]{$x$}
\Text[x=0,y=8]{}
\end{tikzpicture}
\!\!\!\!\!\!\!= {\rm tr}\bigl[\tilde{\mathbb U}^{L}\bigr].
\label{eq:traceid}
\ee
Here the transfer matrix ${\mathbb U}$ corresponds to two consecutive rows, while the `dual transfer matrix' ${\tilde{\mathbb U}\in {\rm End}(\mathcal H_1^{\otimes 2t})}$ corresponds to two consecutive columns, and the boundary conditions in both directions are periodic. As it is clear from the diagram, the dual transfer matrix reads as   
\be
\tilde{\mathbb U}=\mathbb T^{\phantom{\dag}}_{2t} \tilde{U}^{\otimes t} \mathbb T^\dag_{2t} \tilde{U}^{\otimes t}\,,
\label{eq:spacetransfermatrix}
\ee
where we introduced the `dual local gate' $\tilde U$ by means of the following reshuffling
\be
\bra{k}\otimes\bra{\ell}\tilde U \ket{i}\otimes\ket{j}  = \bra{j}\otimes\bra{\ell} U \ket{i}\otimes\ket{k}.
\ee
The dual gate defines the evolution in a circuit where the roles of time and space have been swapped.

In this paper we consider quantum circuits with \emph{unitary} local gates $U$ such that $\tilde U$ \emph{is also unitary}. Namely, we require~\cite{SM}  
\begin{align}
&UU^\dag=U^\dag U = \1 \,\, \Rightarrow \,\,\, \begin{tikzpicture}[baseline=(current  bounding  box.center), scale=0.6]
\draw[thick] (-2.25,1) -- (-1.75,0.5);
\draw[thick] (-1.75,0.5) -- (-1.25,1);
\draw[thick] (-2.25,-1) -- (-1.75,-0.5);
\draw[thick] (-1.75,-0.5) -- (-1.25,-1);
\draw[thick] (-1.9,0.35) to[out=170, in=-170] (-1.9,-0.35);
\draw[thick] (-1.6,0.35) to[out=10, in=-10] (-1.6,-0.35);
\draw[thick, fill=myred, rounded corners=2pt] (-2,0.25) rectangle (-1.5,0.75);
\draw[thick, fill=myblue, rounded corners=2pt] (-2,-0.25) rectangle (-1.5,-0.75);
\Text[x=-0.75,y=0]{$=$}
\draw[thick] (-.25,1) -- (.25,0.5);
\draw[thick] (.25,0.5) -- (.75,1);
\draw[thick] (-.25,-1) -- (.25,-0.5);
\draw[thick] (.25,-0.5) -- (.75,-1);
\draw[thick] (.1,0.35) to[out=170, in=-170] (0.1,-0.35);
\draw[thick] (.4,0.35) to[out=10, in=-10] (0.4,-0.35);
\draw[thick, fill=myblue, rounded corners=2pt] (0,0.25) rectangle (0.5,0.75);
\draw[thick, fill=myred, rounded corners=2pt] (0,-0.25) rectangle (0.5,-0.75);
\Text[x=1.25,y=0]{$=$}
\draw[thick] (2.2,-.75) -- (2.2,.75) (2.8,-0.75) -- (2.8,0.75) (2.2,-.75) -- (1.95,.-1) (2.8,-.75) -- (3.05,.-1) (2.2,.75) -- (1.95,1) (2.8,.75) -- (3.05,1);
\Text[x=3,y=-0.1]{,}
\Text[x=0,y=-1.2]{}
\end{tikzpicture}
\label{eq:unitarity}\\
&\tilde U\tilde U^\dag=\tilde U^\dag \tilde U = \1 \,\, \Rightarrow \, \begin{tikzpicture}[baseline=(current  bounding  box.center), scale=0.6]
\def\eps{0.2};
\draw[thick] (-1.75+2,0.5+\eps) -- (-1.25+2,1+\eps);
\draw[thick] (-1.25+2,0+\eps) -- (-1.75+2,0.5+\eps);
\draw[thick] (-1.25+2,0-\eps) -- (-1.75+2,-0.5-\eps);
\draw[thick] (-1.75+2,-0.5-\eps) -- (-1.25+2,-1-\eps);
\draw[thick] (-1.9+2,0.7+\eps) to[out=170, in=-170] (-1.9+2,-0.7-\eps);
\draw[thick] (-1.9+2,0.3+\eps) to[out=170, in=-170] (-1.9+2,-0.3-\eps);
\draw[thick, fill=myblue, rounded corners=2pt] (-2+2,0.25+\eps) rectangle (-1.5+2,0.75+\eps);
\draw[thick, fill=myred, rounded corners=2pt] (-2+2,-0.25-\eps) rectangle (-1.5+2,-0.75-\eps);
\draw[thick] (-1.95+4,0.7+\eps) to[out=170, in=-170] (-1.95+4,-0.7-\eps);
\draw[thick] (-1.95+4,0.3+\eps) to[out=170, in=-170] (-1.95+4,-0.3-\eps);
\draw[thick] (-1.6+4,0.7+\eps) -- (-1.95+4,0.7+\eps)(-1.6+4,0.7+\eps) -- (-1.25+4,1+\eps);
\draw[thick] (-1.6+4,0.3+\eps) -- (-1.95+4,0.3+\eps)(-1.6+4,0.3+\eps) -- (-1.25+4,+\eps);
\draw[thick] (-1.6+4,-0.3-\eps) -- (-1.95+4,-0.3-\eps)(-1.6+4,-0.3-\eps) -- (-1.25+4,-\eps);
\draw[thick] (-1.6+4,-0.7-\eps) -- (-1.95+4,-0.7-\eps)(-1.6+4,-0.7-\eps) -- (-1.25+4,-1-\eps);
\Text[x=1.15,y=0]{$=$}
\Text[x=2.9,y=-0.05]{,}
\Text[x=0,y=-1.3]{}
\end{tikzpicture}
\,\,
\begin{tikzpicture}[baseline=(current  bounding  box.center), scale=0.6]
\def\eps{0.2};
\draw[thick] (-2.25+2,1+\eps) -- (-1.75+2,0.5+\eps);
\draw[thick] (-1.75+2,0.5+\eps) -- (-2.25+2,0+\eps);
\draw[thick] (-1.75+2,-0.5-\eps) -- (-2.25+2,0-\eps);
\draw[thick] (-2.25+2,-1-\eps) -- (-1.75+2,-0.5-\eps);
\draw[thick] (-1.6+2,0.7+\eps) to[out=10, in=-10] (-1.6+2,-0.7-\eps);
\draw[thick] (-1.6+2,0.3+\eps) to[out=10, in=-10] (-1.6+2,-0.3-\eps);
\draw[thick, fill=myblue, rounded corners=2pt] (-2+2,0.25+\eps) rectangle (-1.5+2,0.75+\eps);
\draw[thick, fill=myred, rounded corners=2pt] (-2+2,-0.25-\eps) rectangle (-1.5+2,-0.75-\eps);
\draw[thick] (-1.6+4,0.7+\eps) to[out=10, in=-10] (-1.6+4,-0.7-\eps);
\draw[thick] (-1.6+4,0.3+\eps) to[out=10, in=-10] (-1.6+4,-0.3-\eps);
\draw[thick] (-1.6+4,0.7+\eps) -- (-1.95+4,0.7+\eps)(-2.25+4,1+\eps) -- (-1.95+4,0.7+\eps);
\draw[thick] (-1.6+4,0.3+\eps) -- (-1.95+4,0.3+\eps)(-2.25+4,+\eps) -- (-1.95+4,0.3+\eps);
\draw[thick] (-1.6+4,-0.3-\eps) -- (-1.95+4,-0.3-\eps)(-2.25+4,-\eps) -- (-1.95+4,-0.3-\eps);
\draw[thick] (-1.6+4,-0.7-\eps) -- (-1.95+4,-0.7-\eps)(-2.25+4,-1-\eps) -- (-1.95+4,-0.7-\eps);
\Text[x=1.35,y=0]{$=$}
\Text[x=3.1,y=-0.05]{.}
\Text[x=0,y=-1.3]{}
\end{tikzpicture}
\label{eq:tildeunitarity}
\end{align}
We call `dual-unitary' local gates fulfilling both \eqref{eq:unitarity} and \eqref{eq:tildeunitarity} (these conditions immediately imply that $\mathbb U$ and $\tilde{\mathbb U}$ are also unitary). In the following we show that dual-unitary gates provide a remarkable testing ground for studying dynamical correlations in many-body quantum systems. They allow us to classify a number of qualitatively different physical behaviours \cite{footnote}.

Here we consider dynamical correlation functions of local operators in the general time-translation invariant, `tracial', or infinite temperature state. These quantities are defined as follows
\be
D^{\alpha\beta}(x,y,t)\equiv \frac{1}{d^{2L}}{\rm tr}\left[a^\alpha_{x} \mathbb U^{-t} a_{y}^\beta \mathbb U^{t} \right]\,,
\label{eq:dynamicalcorrelations}
\ee
where ${x,y\in\frac{1}{2}\mathbb Z_{2L}}$, ${t\in \mathbb N}$ (the space-time lattice of the circuit is drawn in Eq.~\eqref{eq:traceid}) and $\{a_x^{\alpha}\}_{\alpha=0}^{d^2-1}$ denotes a basis of the space of local operators at site $x$, i.e., a basis of ${\rm End}(\mathcal H_1)$.  We assume that $a^\alpha$  are Hilbert-Schmidt 
orthonormal ${\rm tr} \left[ (a^\alpha)^\dag a^\beta\right] = d\, \delta_{\alpha,\beta}$ and choose 
$a^0=\1$, so all other $a^\alpha$ are traceless, i.e., ${\rm tr}[a^\alpha] = 0$ for $\alpha\neq 0$. 

The expression \eqref{eq:dynamicalcorrelations} is represented diagrammatically as 
\begin{equation}
\label{eq:graphrep}
\begin{tikzpicture}[baseline=(current  bounding  box.center), scale=0.45]
\foreach \i in {0,1,2,3}
{
}
\foreach \i in {1,3,5}
{
}
\foreach \i in {0,...,3}
{
}
\foreach \i in {-2,-1}
{
}
\foreach \i in {1,3,5,7}
{
}
\foreach \i in {3,1}
{
}
\foreach \i in {0,...,5}
{
\draw[thick] (-.5-2*\i,1) -- (0.5-2*\i,0);
\draw[thick] (-0.5-2*\i,0) -- (0.5-2*\i,1);
\draw[thick] (-.5-2*\i,-1) -- (0.525-2*\i,0.025);
\draw[thick] (-0.525-2*\i,0.025) -- (0.5-2*\i,-1);
\draw[thick, fill=myblue, rounded corners=2pt] (-0.25-2*\i,0.25) rectangle (.25-2*\i,0.75);
\draw[thick, fill=myred, rounded corners=2pt] (-0.25-2*\i,-0.25) rectangle (.25-2*\i,-0.75);}
\foreach \i in {1,...,6}
{
\draw[thick] (.5-2*\i,6) -- (1-2*\i,5.5);
\draw[thick] (1.5-2*\i,6) -- (1-2*\i,5.5);
\draw[thick] (.5-2*\i,-6) -- (1-2*\i,-5.5);
\draw[thick] (1.5-2*\i,-6) -- (1-2*\i,-5.5);
}
\foreach \jj[evaluate=\jj as \j using -2*(ceil(\jj/2)-\jj/2)] in {0,...,3}
\foreach \i in {1,...,6}
{
\draw[thick] (.5-2*\i-1*\j,2+1*\jj) -- (1-2*\i-1*\j,1.5+\jj);
\draw[thick] (1-2*\i-1*\j,1.5+1*\jj) -- (1.5-2*\i-1*\j,2+\jj);
}
\foreach \jj[evaluate=\jj as \j using -2*(ceil(\jj/2)-\jj/2)] in {0,...,4}
\foreach \i in {1,...,6}
{
\draw[thick] (.5-2*\i-1*\j,1+1*\jj) -- (1-2*\i-1*\j,1.5+\jj);
\draw[thick] (1-2*\i-1*\j,1.5+1*\jj) -- (1.5-2*\i-1*\j,1+\jj);
\draw[thick, fill=myblue, rounded corners=2pt] (0.75-2*\i-1*\j,1.75+\jj) rectangle (1.25-2*\i-1*\j,1.25+\jj);
}
\foreach \jj[evaluate=\jj as \j using -2*(ceil(\jj/2)-\jj/2)] in {0,...,3}
\foreach \i in {1,...,6}
{
\draw[thick] (.5-2*\i-1*\j,-2-1*\jj) -- (1-2*\i-1*\j,-1.5-\jj);
\draw[thick] (1-2*\i-1*\j,-1.5-1*\jj) -- (1.5-2*\i-1*\j,-2-\jj);
}
\foreach \jj[evaluate=\jj as \j using -2*(ceil(\jj/2)-\jj/2)] in {0,...,4}
\foreach \i in {1,...,6}
{
\draw[thick] (.5-2*\i-1*\j,-1-1*\jj) -- (1-2*\i-1*\j,-1.5-\jj);
\draw[thick] (1-2*\i-1*\j,-1.5-1*\jj) -- (1.5-2*\i-1*\j,-1-\jj);
\draw[thick, fill=myred, rounded corners=2pt] (0.75-2*\i-1*\j,-1.75-\jj) rectangle (1.25-2*\i-1*\j,-1.25-\jj);
}
\draw[thick, fill=black] (-6.5,0) circle (0.1cm); 
\draw[thick, fill=black] (-3.5,6) circle (0.1cm);
\Text[x=-7,y=-0.5]{$a^\beta_y$}
\Text[x=-4,y=5.6]{$a^\alpha_x$} 
\Text[x=-12.5, y=0]{$\displaystyle\frac{1}{d^{2L}}$}
\Text[x=1.2, y=0]{,}
\draw [thick, decorate, decoration={brace,amplitude=10pt,mirror,raise=4pt},yshift=0pt]
(0.29,0) -- (0.29,6) node [black,midway,xshift=0.65cm] {$t$};
\end{tikzpicture}
\end{equation}
where, again, boundary conditions in both directions are periodic. 
Since $\mathbb U^{-t} a^0_x \mathbb U^t = a^0_x$, we have for all  $\alpha\neq0$ 
\be
D^{00}(x,y,t)=1,\quad D^{0\alpha}(x,y,t)=D^{\alpha0}(x,y,t)=0.
\ee
Moreover, using the two-site shift invariance of $\mathbb U$, we find  
\be
\!\!\!\!D^{\alpha\beta}(x,y,t)= \!\!
\begin{cases}
C_-^{\alpha\beta}(x-y,t)\,\,\,\,2y\,\,\text{even}\\
C_+^{\alpha\beta}(x-y,t)\,\,\,\,2y\,\,\text{odd}
\end{cases}\!\!,
\ee
where we set  $C_\pm^{\alpha\beta}(x,t)\equiv D^{\alpha\beta}(x+\tfrac{1\mp1}{4},\tfrac{1\mp1}{4},t)$. 

We are now in a position to derive the main result of this letter: an exact closed-form expression for \eqref{eq:dynamicalcorrelations}. The calculation can be subdivided in two main steps, summarised in the following two properties.

\begin{property}
\label{p:prop1}
If $U$ is dual-unitary, the dynamical correlations for $t \le L/2$ are non-zero only on the edges of a lightcone spreading at speed 1
\be
C_\nu^{\alpha\beta}(x,t)= \delta_{x,\nu t} C_\nu^{\alpha\beta}(\nu t,t)\,,\qquad\nu=\pm,\,\,\alpha,\beta\neq0\,.
\ee
\end{property}
Before proceeding with the rigorous proof we note that Property~\ref{p:prop1} has a clear physical interpretation. Due to the dual-unitarity of the dynamics, correlations have a causal cone in space, together with that in time. Since they can only propagate along the intersection of the two lightcones, we must have ${x=\pm t}$.

\begin{proof}
The most intuitive way to prove this property is by using the diagrammatic representation (\ref{eq:graphical},\ref{eq:graphrep}). Let us consider the case $\nu=+$ while the procedure for $\nu=-$ is analogous.  

By repeated use of the unitarity property \eqref{eq:unitarity} we can simplify the circuit (\ref{eq:graphrep}) out of the light-cone spreading at speed $1$ from $a_0^\beta$. This is a simple consequence of the causal structure of the time evolution. Pictorially, we have   
\begin{equation}
C_+^{\alpha\beta}(x,t)\!= 
\frac{1}{d^{4t}}
\begin{tikzpicture}[baseline=(current  bounding  box.center), scale=0.45]
\foreach \j in {0,...,4}{
\draw[thick] (-4.77+\j,1.35+\j) to[out=270, in=-270] (-4.77+\j,-1.35-\j);
\draw[thick] (-7.23-\j,1.35+\j) to[out=-90, in=90] (-7.23-\j,-1.35-\j);}
\draw[thick] (-.5-6,1) -- (-6,0.5);
\draw[thick] (-6,0.5) -- (0.5-6,1);
\draw[thick] (-.5-6,-1) -- (-6,-0.5);
\draw[thick] (-6,-0.5) -- (0.5-6,-1);
\draw[thick] (-.15-6,0.35) to[out=170, in=-170] (-.15-6,-0.35);
\draw[thick] (.15-6,0.35) to[out=10, in=-10] (.15-6,-0.35);
\draw[thick, fill=myblue, rounded corners=2pt] (-0.25-6,0.25) rectangle (.25-6,0.75);
\draw[thick, fill=myred, rounded corners=2pt] (-0.25-6,-0.25) rectangle (.25-6,-0.75);
\foreach \i in {1,...,6}
{
\draw[thick] (.6-2*\i,6) to[out=-90, in=-170] (.85-2*\i,5.65);
\draw[thick] (1.4-2*\i,6) to[out=-90, in=-10] (1.15-2*\i,5.65);
\draw[thick] (.6-2*\i,-6) to[out=90, in=-170] (.85-2*\i,-5.65);
\draw[thick] (1.4-2*\i,-6) to[out=90, in=-10] (1.15-2*\i,-5.65);
}
\foreach \jj[evaluate=\jj as \j using 2*(ceil(\jj/2)-\jj/2), evaluate=\jj as \aa using int(3-\jj/2), evaluate=\jj as \bb using int(4+\jj/2)] in {0,...,3}
\foreach \i in {\aa,...,\bb}
{
\draw[thick] (.5-2*\i-1*\j,2+1*\jj) -- (1-2*\i-1*\j,1.5+\jj);
\draw[thick] (1-2*\i-1*\j,1.5+1*\jj) -- (1.5-2*\i-1*\j,2+\jj);
}
\foreach \jj[evaluate=\jj as \j using 2*(ceil(\jj/2)-\jj/2), evaluate=\jj as \aa using int(3-\jj/2), evaluate=\jj as \bb using int(3+\jj/2)] in {0,...,4}
\foreach \i in {\aa,...,\bb}{
\draw[thick] (.5-2*\i-1*\j,1+1*\jj) -- (1-2*\i-1*\j,1.5+\jj);}
\foreach \jj[evaluate=\jj as \j using 2*(ceil(\jj/2)-\jj/2), evaluate=\jj as \aa using int(4-\jj/2), evaluate=\jj as \bb using int(4+\jj/2)] in {0,...,4}
\foreach \i in {\aa,...,\bb}{
\draw[thick] (1-2*\i-1*\j,1.5+1*\jj) -- (1.5-2*\i-1*\j,1+\jj);
}
\foreach \jj[evaluate=\jj as \j using 2*(ceil(\jj/2)-\jj/2), evaluate=\jj as \aa using int(3-\jj/2), evaluate=\jj as \bb using int(4+\jj/2)] in {0,...,4}
\foreach \i in {\aa,...,\bb}{
\draw[thick, fill=myblue, rounded corners=2pt] (0.75-2*\i-1*\j,1.75+\jj) rectangle (1.25-2*\i-1*\j,1.25+\jj);
}
\foreach \jj[evaluate=\jj as \j using 2*(ceil(\jj/2)-\jj/2), evaluate=\jj as \aa using int(3-\jj/2), evaluate=\jj as \bb using int(4+\jj/2)] in {0,...,3}
\foreach \i in {\aa,...,\bb}
{
\draw[thick] (.5-2*\i-1*\j,-2-1*\jj) -- (1-2*\i-1*\j,-1.5-\jj);
\draw[thick] (1-2*\i-1*\j,-1.5-1*\jj) -- (1.5-2*\i-1*\j,-2-\jj);
}
\foreach \jj[evaluate=\jj as \j using 2*(ceil(\jj/2)-\jj/2), evaluate=\jj as \aa using int(3-\jj/2), evaluate=\jj as \bb using int(3+\jj/2)] in {0,...,4}
\foreach \i in {\aa,...,\bb}{
\draw[thick] (.5-2*\i-1*\j,-1-1*\jj) -- (1-2*\i-1*\j,-1.5-\jj);}
\foreach \jj[evaluate=\jj as \j using 2*(ceil(\jj/2)-\jj/2), evaluate=\jj as \aa using int(4-\jj/2), evaluate=\jj as \bb using int(4+\jj/2)] in {0,...,4}
\foreach \i in {\aa,...,\bb}{
\draw[thick] (1-2*\i-1*\j,-1.5-1*\jj) -- (1.5-2*\i-1*\j,-1-\jj);}
\foreach \jj[evaluate=\jj as \j using 2*(ceil(\jj/2)-\jj/2), evaluate=\jj as \aa using int(3-\jj/2), evaluate=\jj as \bb using int(4+\jj/2)] in {0,...,4}
\foreach \i in {\aa,...,\bb}{
\draw[thick, fill=myred, rounded corners=2pt] (0.75-2*\i-1*\j,-1.75-\jj) rectangle (1.25-2*\i-1*\j,-1.25-\jj);
}
\draw[thick, fill=black] (-6.35,0) circle (0.1cm); 
\draw[thick, fill=black] (-3.4,6) circle (0.1cm);
\Text[x=-6.7,y=0.2]{$a^\beta$}
\Text[x=-3.85,y=6.1]{$a^\alpha$} 
\end{tikzpicture}\,.
\end{equation}
At this point, it is convenient to distinguish three cases: (i) $x=t$; (ii) $x=t-\tfrac{1}{2}$; (iii) $x\neq t-\tfrac{1}{2}, t$. Let us start considering the case (iii), using the unitarity of $\tilde U$, i.e. Eq.~\eqref{eq:tildeunitarity}, we have 
\begin{equation}
\!C_+^{\alpha\beta}(x,t)\!= 
\frac{1}{d^{4t-1}}
\begin{tikzpicture}[baseline=(current  bounding  box.center), scale=0.45]
\foreach \j in {0,...,4}{
\draw[thick] (-4.77+\j,1.35+\j) to[out=270, in=-270] (-4.77+\j,-1.35-\j);
\draw[thick] (-7.23-\j,1.35+\j) to[out=-90, in=90] (-7.23-\j,-1.35-\j);}
\draw[thick] (-.5-6,1) -- (-6,0.5);
\draw[thick] (-6,0.5) -- (0.5-6,1);
\draw[thick] (-.5-6,-1) -- (-6,-0.5);
\draw[thick] (-6,-0.5) -- (0.5-6,-1);
\draw[thick] (-.15-6,0.35) to[out=170, in=-170] (-.15-6,-0.35);
\draw[thick] (.15-6,0.35) to[out=10, in=-10] (.15-6,-0.35);
\draw[thick, fill=myblue, rounded corners=2pt] (-0.25-6,0.25) rectangle (.25-6,0.75);
\draw[thick, fill=myred, rounded corners=2pt] (-0.25-6,-0.25) rectangle (.25-6,-0.75);
\draw[thick] (-1.15,5.35) -- (-.77,5.35);
\draw[thick] (-1.15,-5.35) -- (-.77,-5.35);
\foreach \i in {2,...,6}
{
\draw[thick] (.6-2*\i,6) to[out=-90, in=-170] (.85-2*\i,5.65);
\draw[thick] (1.4-2*\i,6) to[out=-90, in=-10] (1.15-2*\i,5.65);
\draw[thick] (.6-2*\i,-6) to[out=90, in=-170] (.85-2*\i,-5.65);
\draw[thick] (1.4-2*\i,-6) to[out=90, in=-10] (1.15-2*\i,-5.65);
}
\foreach \jj[evaluate=\jj as \j using 2*(ceil(\jj/2)-\jj/2), evaluate=\jj as \aa using int(3-\jj/2), evaluate=\jj as \bb using int(4+\jj/2)] in {0,...,3}
\foreach \i in {\aa,...,\bb}
{
\draw[thick] (.5-2*\i-1*\j,2+1*\jj) -- (1-2*\i-1*\j,1.5+\jj);
\draw[thick] (1-2*\i-1*\j,1.5+1*\jj) -- (1.5-2*\i-1*\j,2+\jj);
}
\foreach \jj[evaluate=\jj as \j using 2*(ceil(\jj/2)-\jj/2), evaluate=\jj as \aa using int(3-\jj/2), evaluate=\jj as \bb using int(3+\jj/2)] in {0,...,4}
\foreach \i in {\aa,...,\bb}{
\draw[thick] (.5-2*\i-1*\j,1+1*\jj) -- (0.85-2*\i-1*\j,1.35+\jj);
}
\foreach \jj[evaluate=\jj as \j using 2*(ceil(\jj/2)-\jj/2), evaluate=\jj as \aa using int(4-\jj/2), evaluate=\jj as \bb using int(4+\jj/2)] in {0,...,4}
\foreach \i in {\aa,...,\bb}{
\draw[thick] (1-2*\i-1*\j,1.5+1*\jj) -- (1.5-2*\i-1*\j,1+\jj);
\draw[thick, fill=myblue, rounded corners=2pt] (0.75-2*\i-1*\j,1.75+\jj) rectangle (1.25-2*\i-1*\j,1.25+\jj);
}
\foreach \i in {1,...,4}{
\draw[thick, fill=myblue, rounded corners=2pt] (-5.75+\i,.75+\i) rectangle (-6.25+\i,.25+\i);
}
\foreach \jj[evaluate=\jj as \j using 2*(ceil(\jj/2)-\jj/2), evaluate=\jj as \aa using int(3-\jj/2), evaluate=\jj as \bb using int(4+\jj/2)] in {0,...,3}
\foreach \i in {\aa,...,\bb}
{
\draw[thick] (.5-2*\i-1*\j,-2-1*\jj) -- (1-2*\i-1*\j,-1.5-\jj);
\draw[thick] (1-2*\i-1*\j,-1.5-1*\jj) -- (1.5-2*\i-1*\j,-2-\jj);
}
\foreach \jj[evaluate=\jj as \j using 2*(ceil(\jj/2)-\jj/2), evaluate=\jj as \aa using int(3-\jj/2), evaluate=\jj as \bb using int(3+\jj/2)] in {0,...,4}
\foreach \i in {\aa,...,\bb}{
\draw[thick] (.5-2*\i-1*\j,-1-1*\jj) -- (0.85-2*\i-1*\j,-1.35-\jj);}
\foreach \jj[evaluate=\jj as \j using 2*(ceil(\jj/2)-\jj/2), evaluate=\jj as \aa using int(4-\jj/2), evaluate=\jj as \bb using int(4+\jj/2)] in {0,...,4}
\foreach \i in {\aa,...,\bb}{
\draw[thick] (1-2*\i-1*\j,-1.5-1*\jj) -- (1.5-2*\i-1*\j,-1-\jj);
\draw[thick, fill=myred, rounded corners=2pt] (0.75-2*\i-1*\j,-1.75-\jj) rectangle (1.25-2*\i-1*\j,-1.25-\jj);
}
\foreach \i in {1,...,4}{
\draw[thick, fill=myred, rounded corners=2pt] (-5.75+\i,-.75-\i) rectangle (-6.25+\i,-.25-\i);
}
\draw[thick, fill=black] (-6.35,0) circle (0.1cm); 
\draw[thick, fill=black] (-3.4,6) circle (0.1cm);
\Text[x=-6.7,y=0.2]{$a^\beta$}
\Text[x=-3.85,y=6.1]{$a^\alpha$} 
\end{tikzpicture}\,.
\end{equation}
From this picture it is clear that \eqref{eq:tildeunitarity} can be `telescoped' until the operator $a^\beta$ is encountered. Namely  
\begin{equation}
C_+^{\alpha\beta}(x,t)\!=\! 
\frac{1}{d^{4t-1}}
\begin{tikzpicture}[baseline=(current  bounding  box.center), scale=0.45]
\foreach \j in {0,...,4}{
\draw[thick] (-4.85+\j,1.35+\j) to[out=270, in=-270] (-4.85+\j,-1.35-\j);
\draw[thick] (-7.23-\j,1.35+\j) to[out=-90, in=90] (-7.23-\j,-1.35-\j);}
\draw[thick] (-.5-6,1) -- (-6.15,0.65);
\draw[thick] (-5.85,0.65) -- (0.5-6,1);
\draw[thick] (-.5-6,-1) -- (-6.15,-0.65);
\draw[thick] (-5.85,-0.65) -- (0.5-6,-1);
\draw[thick] (-.15-6,0.35) to[out=170, in=-170] (-.15-6,-0.35);
\draw[thick] (.15-6,0.35) to[out=10, in=-10] (.15-6,-0.35);
\draw[thick] (-1.15,5.35) -- (-.85,5.35);
\draw[thick] (-1.15,-5.35) -- (-.85,-5.35);
\draw[thick] (-.15-6,.35) -- (.15-6,.35);
\draw[thick] (-.15-6,-.35) -- (.15-6,-.35);
\draw[thick] (-.15-6,.65) -- (.15-6,.65);
\draw[thick] (-.15-6,-.65) -- (.15-6,-.65);
\foreach \i in {2,...,6}
{
\draw[thick] (.6-2*\i,6) to[out=-90, in=-170] (.85-2*\i,5.65);
\draw[thick] (1.4-2*\i,6) to[out=-90, in=-10] (1.15-2*\i,5.65);
\draw[thick] (.6-2*\i,-6) to[out=90, in=-170] (.85-2*\i,-5.65);
\draw[thick] (1.4-2*\i,-6) to[out=90, in=-10] (1.15-2*\i,-5.65);
}
\foreach \jj[evaluate=\jj as \j using 2*(ceil(\jj/2)-\jj/2), evaluate=\jj as \aa using int(3-\jj/2), evaluate=\jj as \bb using int(4+\jj/2)] in {0,...,3}
\foreach \i in {\aa,...,\bb}
{
\draw[thick] (.5-2*\i-1*\j,2+1*\jj) -- (0.85-2*\i-1*\j,1.65+\jj);
\draw[thick] (1.15-2*\i-1*\j,1.65+1*\jj) -- (1.5-2*\i-1*\j,2+\jj);
\draw[thick] (1.15-2*\i-1*\j,1.65+1*\jj) -- (.85-2*\i-1*\j,1.65+1*\jj);
\draw[thick] (1.15-2*\i-1*\j,1.35+1*\jj) -- (.85-2*\i-1*\j,1.35+1*\jj);
}
\foreach \jj[evaluate=\jj as \j using 2*(ceil(\jj/2)-\jj/2), evaluate=\jj as \aa using int(3-\jj/2), evaluate=\jj as \bb using int(3+\jj/2)] in {0,...,4}
\foreach \i in {\aa,...,\bb}{
\draw[thick] (.5-2*\i-1*\j,1+1*\jj) -- (0.85-2*\i-1*\j,1.35+\jj);
}
\foreach \jj[evaluate=\jj as \j using 2*(ceil(\jj/2)-\jj/2), evaluate=\jj as \aa using int(4-\jj/2), evaluate=\jj as \bb using int(4+\jj/2)] in {0,...,4}
\foreach \i in {\aa,...,\bb}{
\draw[thick] (1-2*\i-1*\j,1.5+1*\jj) -- (1.5-2*\i-1*\j,1+\jj);
\draw[thick, fill=myblue, rounded corners=2pt] (0.75-2*\i-1*\j,1.75+\jj) rectangle (1.25-2*\i-1*\j,1.25+\jj);
}
\foreach \jj[evaluate=\jj as \j using 2*(ceil(\jj/2)-\jj/2), evaluate=\jj as \aa using int(3-\jj/2), evaluate=\jj as \bb using int(4+\jj/2)] in {0,...,3}
\foreach \i in {\aa,...,\bb}
{
\draw[thick] (.5-2*\i-1*\j,-2-1*\jj) -- (0.85-2*\i-1*\j,-1.65-\jj);
\draw[thick] (1.15-2*\i-1*\j,-1.65-1*\jj) -- (1.5-2*\i-1*\j,-2-\jj);
\draw[thick] (1.15-2*\i-1*\j,-1.65-1*\jj) -- (.85-2*\i-1*\j,-1.65-1*\jj);
\draw[thick] (1.15-2*\i-1*\j,-1.35-1*\jj) -- (.85-2*\i-1*\j,-1.35-1*\jj);
}
\foreach \jj[evaluate=\jj as \j using 2*(ceil(\jj/2)-\jj/2), evaluate=\jj as \aa using int(3-\jj/2), evaluate=\jj as \bb using int(3+\jj/2)] in {0,...,4}
\foreach \i in {\aa,...,\bb}{
\draw[thick] (.5-2*\i-1*\j,-1-1*\jj) -- (0.85-2*\i-1*\j,-1.35-\jj);}
\foreach \jj[evaluate=\jj as \j using 2*(ceil(\jj/2)-\jj/2), evaluate=\jj as \aa using int(4-\jj/2), evaluate=\jj as \bb using int(4+\jj/2)] in {0,...,4}
\foreach \i in {\aa,...,\bb}{
\draw[thick] (1-2*\i-1*\j,-1.5-1*\jj) -- (1.5-2*\i-1*\j,-1-\jj);
\draw[thick, fill=myred, rounded corners=2pt] (0.75-2*\i-1*\j,-1.75-\jj) rectangle (1.25-2*\i-1*\j,-1.25-\jj);
}
\draw[thick, fill=black] (-6.35,0) circle (0.1cm); 
\draw[thick, fill=black] (-3.4,6) circle (0.1cm);
\Text[x=-6.7,y=0.2]{$a^\beta$}
\Text[x=-3.85,y=6.1]{$a^\alpha$} 
\end{tikzpicture}\,,
\end{equation}
where the central loop represents the trace of $a^\beta$ factoring out. Using that for $\beta\neq0$ the operators $a^\beta$ are traceless, we then conclude that the correlation vanishes. 

Consider now the case (ii). Using \eqref{eq:tildeunitarity} we find   
\begin{equation}
\!\!C_+^{\alpha\beta}\!\!\left(t-\tfrac{1}{2},t\right)\!\!=\! 
\frac{1}{d^{4t}}\!
\begin{tikzpicture}[baseline=(current  bounding  box.center), scale=0.45]
\foreach \j in {0,...,4}{
\draw[thick] (-4.77+\j,1.35+\j) to[out=270, in=-270] (-4.77+\j,-1.35-\j);
\draw[thick] (-7.23-\j,1.35+\j) to[out=-90, in=90] (-7.23-\j,-1.35-\j);}
\draw[thick] (-.5-6,1) -- (-6,0.5);
\draw[thick] (-6,0.5) -- (0.5-6,1);
\draw[thick] (-.5-6,-1) -- (-6,-0.5);
\draw[thick] (-6,-0.5) -- (0.5-6,-1);
\draw[thick] (-.15-6,0.35) to[out=170, in=-170] (-.15-6,-0.35);
\draw[thick] (.15-6,0.35) to[out=10, in=-10] (.15-6,-0.35);
\draw[thick, fill=myblue, rounded corners=2pt] (-0.25-6,0.25) rectangle (.25-6,0.75);
\draw[thick, fill=myred, rounded corners=2pt] (-0.25-6,-0.25) rectangle (.25-6,-0.75);
\draw[thick] (-1.15,5.35) -- (-.77,5.35);
\draw[thick] (-1.15,-5.35) -- (-.77,-5.35);
\draw[thick] (-1.15,5.65) -- (-.85,5.65);
\draw[thick] (-1.15,-5.65) -- (-.85,-5.65);
\foreach \i in {1,...,6}
{
\draw[thick] (.6-2*\i,6) to[out=-90, in=-170] (.85-2*\i,5.65);
\draw[thick] (1.4-2*\i,6) to[out=-90, in=-10] (1.15-2*\i,5.65);
\draw[thick] (.6-2*\i,-6) to[out=90, in=-170] (.85-2*\i,-5.65);
\draw[thick] (1.4-2*\i,-6) to[out=90, in=-10] (1.15-2*\i,-5.65);
}
\foreach \jj[evaluate=\jj as \j using 2*(ceil(\jj/2)-\jj/2), evaluate=\jj as \aa using int(3-\jj/2), evaluate=\jj as \bb using int(4+\jj/2)] in {0,...,3}
\foreach \i in {\aa,...,\bb}
{
\draw[thick] (.5-2*\i-1*\j,2+1*\jj) -- (1-2*\i-1*\j,1.5+\jj);
\draw[thick] (1-2*\i-1*\j,1.5+1*\jj) -- (1.5-2*\i-1*\j,2+\jj);
}
\foreach \jj[evaluate=\jj as \j using 2*(ceil(\jj/2)-\jj/2), evaluate=\jj as \aa using int(3-\jj/2), evaluate=\jj as \bb using int(3+\jj/2)] in {0,...,4}
\foreach \i in {\aa,...,\bb}{
\draw[thick] (.5-2*\i-1*\j,1+1*\jj) -- (0.85-2*\i-1*\j,1.35+\jj);
}
\foreach \jj[evaluate=\jj as \j using 2*(ceil(\jj/2)-\jj/2), evaluate=\jj as \aa using int(4-\jj/2), evaluate=\jj as \bb using int(4+\jj/2)] in {0,...,4}
\foreach \i in {\aa,...,\bb}{
\draw[thick] (1-2*\i-1*\j,1.5+1*\jj) -- (1.5-2*\i-1*\j,1+\jj);
\draw[thick, fill=myblue, rounded corners=2pt] (0.75-2*\i-1*\j,1.75+\jj) rectangle (1.25-2*\i-1*\j,1.25+\jj);
}
\foreach \i in {1,...,4}{
\draw[thick, fill=myblue, rounded corners=2pt] (-5.75+\i,.75+\i) rectangle (-6.25+\i,.25+\i);
}
\foreach \jj[evaluate=\jj as \j using 2*(ceil(\jj/2)-\jj/2), evaluate=\jj as \aa using int(3-\jj/2), evaluate=\jj as \bb using int(4+\jj/2)] in {0,...,3}
\foreach \i in {\aa,...,\bb}
{
\draw[thick] (.5-2*\i-1*\j,-2-1*\jj) -- (1-2*\i-1*\j,-1.5-\jj);
\draw[thick] (1-2*\i-1*\j,-1.5-1*\jj) -- (1.5-2*\i-1*\j,-2-\jj);
}
\foreach \jj[evaluate=\jj as \j using 2*(ceil(\jj/2)-\jj/2), evaluate=\jj as \aa using int(3-\jj/2), evaluate=\jj as \bb using int(3+\jj/2)] in {0,...,4}
\foreach \i in {\aa,...,\bb}{
\draw[thick] (.5-2*\i-1*\j,-1-1*\jj) -- (0.85-2*\i-1*\j,-1.35-\jj);}
\foreach \jj[evaluate=\jj as \j using 2*(ceil(\jj/2)-\jj/2), evaluate=\jj as \aa using int(4-\jj/2), evaluate=\jj as \bb using int(4+\jj/2)] in {0,...,4}
\foreach \i in {\aa,...,\bb}{
\draw[thick] (1-2*\i-1*\j,-1.5-1*\jj) -- (1.5-2*\i-1*\j,-1-\jj);
\draw[thick, fill=myred, rounded corners=2pt] (0.75-2*\i-1*\j,-1.75-\jj) rectangle (1.25-2*\i-1*\j,-1.25-\jj);
}
\foreach \i in {1,...,4}{
\draw[thick, fill=myred, rounded corners=2pt] (-5.75+\i,-.75-\i) rectangle (-6.25+\i,-.25-\i);
}
\draw[thick, fill=black] (-6.35,0) circle (0.1cm); 
\draw[thick, fill=black] (-1.4,6) circle (0.1cm);
\Text[x=-6.75,y=0.2]{$a^\beta$}
\Text[x=-1.85,y=6.2]{$a^\alpha$} 
\end{tikzpicture}.
\end{equation}
Here the loop giving ${\rm tr}[a^\alpha]$ factors out so we again conclude that the whole expression vanishes. We then showed that the only remaining possibility is the case (i). This concludes the proof. 
\end{proof}

\begin{property}
\label{p:prop2}
The light cone correlations $C_+^{\alpha\beta}(t,t)$ and $C_-^{\alpha\beta}(-t,t)$ are given by  
\begin{align}
C_\nu^{\alpha\beta}(\nu t,t) &= \frac{1}{d} {\rm tr}\left[\mathcal M_{\nu}^{2t}(a^\beta)a^\alpha\right]\!,\label{eq:C+}
\end{align}
where we introduced the linear maps over ${\rm End}(\mathbb C^d)$
\begin{align}
&\mathcal M_{+}(a) = \frac{1}{d} {\rm tr}_1\left[U^\dag (a\otimes\1) U\right]=
\frac{1}{d} 
\begin{tikzpicture}[baseline=(current  bounding  box.center), scale=0.7]
\draw[thick] (-6.7,.7) -- (-6.1,.7);
\draw[thick] (-6.7,-.7) -- (-6.1,-.7);
\draw[thick] (-5.8,0.5) -- (-5.8,1);
\draw[thick] (-5.8,-0.5) -- (-5.8,-1);
\draw[thick] (-.7-6,0.7) to[out=210, in=-210] (-.7-6,-0.7);
\draw[thick] (-.15-6,0.35) to[out=170, in=-170] (-.15-6,-0.35);
\draw[thick] (.15-6,0.35) to[out=10, in=-10] (.15-6,-0.35);
\draw[thick, fill=myblue, rounded corners=2pt] (-0.25-6,0.25) rectangle (.25-6,0.75);
\draw[thick, fill=myred, rounded corners=2pt] (-0.25-6,-0.25) rectangle (.25-6,-0.75);
\draw[thick, fill=black] (-6.35,0) circle (0.1cm); 
\Text[x=-6.7,y=0.1]{$a$}
\end{tikzpicture},
\label{eq:defrho+}\\
&\mathcal M_{-}(a) = \frac{1}{d} {\rm tr}_2\left[U^\dag (\1\otimes a) U\right]=\frac{1}{d} 
\begin{tikzpicture}[baseline=(current  bounding  box.center), scale=0.7]
\draw[thick] (-5.8,.7) -- (-5.2,.7);
\draw[thick] (-5.8,-.7) -- (-5.2,-.7);
\draw[thick] (-6.2,0.5) -- (-6.2,1);
\draw[thick] (-6.2,-0.5) -- (-6.2,-1);
\draw[thick] (-5.2,0.7) to[out=180-210, in=180+210] (-5.2,-0.7);
\draw[thick] (-.15-6,0.35) to[out=170, in=-170] (-.15-6,-0.35);
\draw[thick] (.15-6,0.35) to[out=10, in=-10] (.15-6,-0.35);
\draw[thick, fill=myblue, rounded corners=2pt] (-0.25-6,0.25) rectangle (.25-6,0.75);
\draw[thick, fill=myred, rounded corners=2pt] (-0.25-6,-0.25) rectangle (.25-6,-0.75);
\draw[thick, fill=black] (-5.65,0) circle (0.1cm); 
\Text[x=-5.3,y=0.1]{$a$}
\end{tikzpicture}.
\label{eq:defrho-}
\end{align}
${\rm tr}_{i}[A]$ denote partial traces over $i$-th site ($i=1,2$).
\end{property}
\begin{proof}
We again prove the property for $C_+^{\alpha\beta}(t,t)$, using the diagrammatic representation. A completely analogous reasoning applies for $C_-^{\alpha\beta}(-t,t)$. 

By repeated use of the unitarity property \eqref{eq:unitarity} we can reduce $C_+^{\alpha\beta}(t,t)$ to the following form
\begin{equation}
\!\!\!\!\!C_+^{\alpha\beta}\!(t,t)\!\!=\!\frac{1}{d^{2t+1}}
\begin{tikzpicture}[baseline=(current  bounding  box.center), scale=0.45]
\foreach \j in {0,...,4}{
\draw[thick] (-4.77+\j,1.32+\j) -- (-4.77+\j,-1.32-\j);
\draw[thick] (-7.5-\j,1.28+\j) -- (-7.5-\j,-1.28-\j);}
\foreach \j in {0,...,4}{
\draw[thick] (-7-\j,1.73+\j) -- (-5.1+\j,1.73+\j);
\draw[thick] (-7-\j,-1.73-\j) -- (-5.1+\j,-1.73-\j);
}
\draw[thick] (-7,.7) -- (-6.1,.7);
\draw[thick] (-7,-.7) -- (-6.1,-.7);
\draw[thick] (-.25,6.25) -- (-0.25,-6.25);
\draw[thick] (-0.5,-6) -- (-.25,-6.25);
\draw[thick] (-0.5,6) -- (-.25,6.25);
\draw[thick] (-6,0.5) -- (0.5-6,1);
\draw[thick] (-6,-0.5) -- (0.5-6,-1);
\draw[thick] (-1-6,0.7) to[out=230, in=-230] (-1-6,-0.7);
\draw[thick] (-.15-6,0.35) to[out=170, in=-170] (-.15-6,-0.35);
\draw[thick] (.15-6,0.35) to[out=10, in=-10] (.15-6,-0.35);
\draw[thick, fill=myblue, rounded corners=2pt] (-0.25-6,0.25) rectangle (.25-6,0.75);
\draw[thick, fill=myred, rounded corners=2pt] (-0.25-6,-0.25) rectangle (.25-6,-0.75);
\foreach \i in {1,...,5}{
\draw[thick] (-6.5-\i,0.25+\i) -- (-6-\i,0.75+\i);
\draw[thick] (-6.5-\i,-0.25-\i) -- (-6-\i,-0.75-\i);
\draw[thick] (-6.5+\i,\i) -- (-5.5+\i,1+\i);
\draw[thick] (-6.5+\i,-\i) -- (-5.5+\i,-1-\i);
\draw[thick, fill=myblue, rounded corners=2pt] (-5.75+\i,.75+\i) rectangle (-6.25+\i,.25+\i);
\draw[thick, fill=myred, rounded corners=2pt] (-5.75+\i,-.75-\i) rectangle (-6.25+\i,-.25-\i);
}
\draw[thick, fill=black] (-6.35,0) circle (0.1cm); 
\draw[thick, fill=black] (-0.25,6.25) circle (0.1cm);
\Text[x=-6.75,y=0.2]{$a^{\beta}$}
\Text[x=-0.8,y=6.4]{$a^\alpha$} 
\end{tikzpicture}.
\label{eq:CPTPdiagram}
\end{equation}
Using the definition \eqref{eq:defrho+} we see that \eqref{eq:CPTPdiagram} is precisely the diagrammatic representation of \eqref{eq:C+}. 
\end{proof}
 Properties~\ref{p:prop1} and \ref{p:prop2} have a very powerful consequence: all dynamical correlations of local operators in dual-unitary quantum circuits are determined by the two linear single-qudit channels $\mathcal M_\pm$. These maps are \emph{trace preserving}, \emph{completely positive}, and \emph{unital} (meaning that they map the identity operator to itself). Moreover, as it is apparent from their definition, they are completely determined by a single $d^2\times d^2$ unitary matrix ($U$ in our case). Maps with these properties are known in the literature as \emph{unistochastic maps}~\cite{ZB:quantummaps, ZB:book, MKZ:unistochastic}. These maps are generically non-diagonalisable, however, they are \emph{contractive}. Namely, the eigenvalues $\{\lambda_{\nu,\gamma}\}_{\gamma=0}^{d^2-1}$ of ${\cal M}_\nu$ lie on the unit disk and those that are on the unit circle have coinciding algebraic and geometric multiplicity~\cite{SM}.  This means that the dynamical correlations take the following general form
\be
C^{\alpha\beta}_\nu(x,t) = \delta_{\nu x,t}  \sum_{\gamma=1}^{d^2-1} c^{\alpha\beta}_{\nu,\gamma} (\lambda_{\nu,\gamma})^{2t},\quad
\alpha,\beta\neq 0\,,
\label{eq:cf}
\ee
where $|\lambda_{\nu,\gamma}|\leq1$ and for eigenvalues corresponding to nontrivial Jordan blocks, the `constant' $c^{\alpha\beta}_{\nu,\gamma}$ is a polynomial in $t$.
Note that since $a^\beta$ are orthogonal we excluded the trivial eigenvalues ${\lambda_{\pm,0}=1}$ corresponding to the identity operator.

This gives a systematic way to classify dual-unitary circuits based on the increasing level of ergodicity of ultra-local observables:
\begin{enumerate}
\item[(i)] Non-interacting behaviour: all $2(d^2-1)$ nontrivial eigenvalues $\lambda_{\nu,\gamma}$ are equal to 1, meaning that all dynamical correlations remain constant.
\item[(ii)]  Non-ergodic (and generically interacting and {\em non-integrable}) behaviour: There are $n$, $1 \le n < 2(d^2-1)$, nontrivial eigenvalues $\lambda_{\nu,\gamma}$ equal to 1, meaning that some dynamical correlations remain constant.
\item[(iii)]  Ergodic but non-mixing behaviour: all nontrivial eigenvalues $\lambda_{\nu,\gamma}$ are different from 1, but there exists at least one eigenvalue with unit modulus. In this case, all time averaged dynamical correlations vanish at large times, reproducing the infinite-temperature state value.
\item[(iv)]  Ergodic and mixing behaviour: all nontrivial eigenvalues are within unit disk, $|\lambda_{\nu,\gamma}| < 1$. In this case, all time dynamical correlations vanish at large times, reproducing the infinite-temperature state value even without time averaging.
\end{enumerate}
An example of (i) is the SWAP gate $U \ket{i}\otimes\ket{j} = \ket{j}\otimes \ket{i}$,
which is clearly self dual, i.e. $U=\tilde{U}$. Note that, since dual-unitary gates are generically not parity invariant, we can have `chiral' cases where the number of non-decaying modes (i.e. with $\lambda_{\nu,\gamma}=e^{i \theta}$) propagating to the left and to the right is different.   

We point out that Eq.~\eqref{eq:cf} gives direct access to time correlations among extensive operators of the form ${A_\nu^\alpha\equiv\sum_{x\in \mathbb Z_L} a_{x+\frac{(\nu-1)}{4}}}$. Specifically one finds 
\be
{\frac{1}{L d^{2L}}{\rm tr}\left[A_\nu^\alpha \mathbb U^{-t} A_\mu^\beta \mathbb{U}^t\right]\! = \! \delta_{\nu,\mu}\sum_{\gamma= 1}^{d^2-1} c^{\alpha\beta}_{\nu,\gamma} (\lambda_{\nu,\gamma})^{2t}}.
\label{eq:timecorr}
\ee 
These correlations are able to distinguish dynamical mixing from the mere decaying local correlators. Indeed, even if all dynamical correlations \eqref{eq:cf} (at fixed distance $x$) vanish in the infinite time limit, the correlation \eqref{eq:timecorr} vanishes only if the system is ergodic and mixing (all nontrivial eigenvalues are within unit disk). In particular, if the mode $a^\alpha_x$ is conserved (i.e. ${\mathcal M_\nu(a^\alpha)=a^\alpha}$), $A_\nu^\alpha$ is a proper conserved charge of the system. Note that, since we considered modes localised on a single site, these conservation laws  are of `single-body form', i.e., they do not couple different sites. Finally, we remark that our classification here only concerns ergodicity 
of ultra-local observables and their extensive sums. 
For instance, circuits in (iv) may in principle still exhibit non-ergodic or non-mixing behaviour for non-local operators or local operators with a larger support.

The proposed classification can be explicitly carried out for $d=2$. Indeed, in this minimal case it is possible to parametrise all dual-unitary local gates~\cite{SM}. The result reads as  
\be
U=e^{i \phi} (u_+ \otimes u_-)\cdot V[J]\cdot (v_-\otimes v_+)\,,
\label{eq:dualunitaryU}
\ee
where $\phi, J \in \mathbb R$~\cite{note2}, $u_\pm,v_\pm\in {\rm SU}(2)$ and 
\be
\!\!\!\!V[J]\!=\! \exp\!\!\left[-i\!\left(\frac{\pi}{4} \sigma^x\otimes\sigma^x \!+\! \frac{\pi}{4} \sigma^y\otimes\sigma^y\!+\! J \sigma^z\otimes\sigma^z\right)\right]\!.
\ee
Quantum circuits with local gates of the form \eqref{eq:dualunitaryU} include both integrable~\cite{PoGri17} and non-integrable cases. For instance, $U_{\rm XXZ}=V[J]$ is a full parameter line of the integrable trotterized XXZ chain~\cite{Lenart1, Lenart2} and 
\be
\!\!\!U_{\rm SDKI}= \!e^{-i h \sigma^{z}} \!\!e^{i \frac{\pi}{4} \sigma^{x}}\!\!\!\otimes e^{i \frac{\pi}{4} \sigma^{x}}\cdot \tilde V[0] \cdot e^{-i h \sigma^{z}}\!\!\!\otimes \1, 
\label{eq:SDKI}
\ee
with $\tilde V[0]= e^{-i \frac{\pi}{4} \sigma^{y}}\!\otimes e^{-i \frac{\pi}{4} \sigma^{y}}\cdot V[0] \cdot e^{i \frac{\pi}{4} \sigma^{y}}\!\!\otimes e^{i \frac{\pi}{4} \sigma^{y}}$, is a quantum circuit representation~\cite{SM} (see also \cite{austen}) of the non-integrable self-dual kicked Ising (SDKI) chain studied in Ref.~\cite{letter,entropy}. In other words, for ${d=2}$ integrable~\cite{PoGri17} and dual-unitary quantum circuits form two distinct but overlapping sets.

Plugging the form \eqref{eq:dualunitaryU} in the definitions \eqref{eq:defrho+} and \eqref{eq:defrho-}, and writing the corresponding matrices in the Pauli basis $\{a^0,a^1, a^2, a^3\}=\{\1,\sigma^x,\sigma^y, \sigma^z\}$ we find 
\begin{align}
\!\!\!\mathcal M_{\pm} &= 
\begin{bmatrix}
1 & 0 \\
0 & R[u_\pm]
\end{bmatrix}\cdot
\mathcal M[J]
\cdot
\begin{bmatrix}
1 & 0 \\
0 & R[v_\pm]
\end{bmatrix}\,.
\end{align}
Here, to lighten notation, we denoted the matrices associated with $\mathcal M_{\pm}$ with the same symbols. Moreover, we denoted by $R[w]$ the $3\times 3$ adjoint representation of $w \in{\rm SU}(2)$, and, finally, we introduced ${\mathcal M[J]\equiv {\rm diag}(1,\sin(2J),\sin(2J),1)}$, the matrix of the map associated with the gate $V[J]$. 

Since the spectrum is invariant under similarity transformations, the eigenvalues of $\mathcal M_{\pm}$ depend only on $J$ and on the products ${v_\pm u_\pm \in{\rm SU}(2)}$, thus, in principle, on four real parameters. The matrix $\mathcal M[J]$, however, is invariant under rotations around the $z$ axis, so the eigenvalues of $\mathcal M_{\pm}$ depend on three real parameters only ($J$ is a common parameter). Analysing the spectra of $\mathcal M_{\pm}$ as functions of the parameters we identify all four types of ergodic behaviour~\cite{SM}. For instance, for the SDKI chain the spectra of $\mathcal M_{+}$ and $\mathcal M_{-}$ coincide and are given by $\{1,\cos(2h),0\}$. This means that, for generic $h$, the model is in the ergodic and mixing class, while at the integrable point $h=0$ it is in the class (ii). At this special point, the $y$-magnetisations on the integer and half-odd integer sublattices are conserved. Instead, in the case of the trotterised XXZ chain the matrices $\mathcal M_{\pm}$ coincide with $\mathcal M[J]$. Therefore, they are always in the class (ii) except for $J={\pi}/{4}$, when they correspond to the SWAP gate and are in the non-interacting class (i). For $J\neq\pi/4$ the charges associated to the conserved modes are $z$-magnetisations on the integer and half-odd integer sublattices.  

The results presented in this letter admit several generalisations and extensions. First of all we note that the proofs of Properties \ref{p:prop1} and \ref{p:prop2} do not rely on translational invariance neither in time nor in space. This means that \eqref{eq:C+} directly generalises to cases where some inhomogeneity or randomness is introduced either in space or in time~\cite{RandomCircuitsEnt, Keyserlingk, Chalker}: one simply needs to replace $\mathcal M^{2t}_\pm(a)$ in \eqref{eq:C+} with a product of $2t$ different operator maps, each one determined by a different local gate depending on the space-time point. 
Moreover, our treatment can be straightforwardly repeated to find correlation functions of local observables with larger support. This will, for instance, allow one to find exactly solvable circuits with more complex, ``many-body'', local conservation laws. Such circuits are currently attracting considerable attention, see e.g. Refs.~\cite{RPK:conservationlaws, H:conservationlaws},  because they are regarded as toy models for generic closed systems. Another very interesting direction is to approach generic quantum circuits by a perturbative expansion around ergodic and mixing dual-unitary ones. Indeed, the fact that in the dual-unitary `point' the dynamics have a sort of exponential space-time clustering hints that an expansion might have a finite radius of convergence. Finally, a construction very similar to the one presented here can be carried out for higher dimensional quantum circuits, where, instead of on chains, one considers local sites disposed on hypercubes of any dimension. Requiring dual-unitarity in all directions again constrains the correlations to the edges of a light cone, and allows one to express them in terms of unistochastic maps.  

\emph{Acknowledgments.} We thank Marek Gluza and Marko \v Znidari\v c for useful discussions. The work has been supported by ERC Advanced grant 694544 -- OMNES and the program P1-0402 of Slovenian Research Agency.

\onecolumngrid

\pagebreak

\newcounter{equationSM}
\newcounter{figureSM}
\newcounter{tableSM}
\stepcounter{equationSM}
\setcounter{equation}{0}
\setcounter{figure}{0}
\setcounter{table}{0}
\makeatletter
\renewcommand{\theequation}{\textsc{sm}-\arabic{equation}}
\renewcommand{\thefigure}{\textsc{sm}-\arabic{figure}}
\renewcommand{\thetable}{\textsc{sm}-\arabic{table}}

\begin{center}
{\large{\bf Supplemental Material for\\
 ``Exact Correlation Functions for Dual-Unitary Lattice Models in 1+1 Dimensions''}}
\end{center}

Here we report some useful information complementing the main text. In particular
\begin{itemize}
\item[-] In Section \ref{sec:FR} we report a detailed derivation of the diagrammatic fusion rules (6) and (7);
\item[-] In Section \ref{sec:mapsdetails} we prove the contracting nature of $\mathcal M_\pm$;
\item[-] In Section \ref{app:dualunitarymatrix} we derive a simple parametrisation of all ${4\times4}$ dual-unitary matrices;
\item[-] In Section \ref{app:kickedising} we derive a quantum unitary circuit formulation of the self-dual kicked Ising model;
\item[-] In Section \ref{app:polynomial} we identify all possible occurrences of unimodular eigenvalues of $\mathcal M_\pm$ for $d=2$;
\end{itemize}

\section{Duality and fusion rules}
\label{sec:FR}

In this section we derive the diagrammatic fusion rules (6) and (7). Let us begin by writing the matrix elements of the quantum gate as 
\be
U_{ij}^{kl} = \bra{k}\otimes \bra{l} U \ket{i}\otimes \ket{j}\qquad\Rightarrow\qquad
\begin{tikzpicture}[baseline=(current  bounding  box.center), scale=0.8]
\def\eps{0.3};
\def\ep{0.65};
\draw[thick] (-4.25,0.5) -- (-3.25,-0.5);
\draw[thick] (-4.25,-0.5) -- (-3.25,0.5);
\draw[thick, fill=myred, rounded corners=2pt] (-4,0.25) rectangle (-3.5,-0.25);
\Text[x=-4.25,y=-0.5-\eps]{$i$}
\Text[x=-4.25,y=-0.5-\ep]{}
\Text[x=-3.25,y=-0.5-\eps]{$j$}
\Text[x=-4.25,y=+0.5+\eps]{$k$}
\Text[x=-3.25,y=+0.5+\eps]{$l$}
\end{tikzpicture},
\ee
then the duality mapping amounts to a Choi-Jamio\l{}kowski reshuffling~\cite{ZB:quantummaps, ZB:book, MKZ:unistochastic}
\be
\tilde{U}_{ik}^{jl} = U_{ij}^{kl}.
\ee
The direct fusion rules are the straightforward tensor network expression of the unitarity of $U$:
\begin{eqnarray}
\sum_{p,q} (U^\dag)_{pq}^{k\ell} U_{ij}^{pq} &=& (U^\dag U)_{ij}^{k\ell} =  \delta_{ik} \delta_{j\ell} \qquad\Rightarrow\qquad 
\begin{tikzpicture}[baseline=(current  bounding  box.center), scale=0.6]
\def\ep{1.3};
\def\eps{0.8};
\draw[thick] (-2.25,1) -- (-1.75,0.5);
\draw[thick] (-1.75,0.5) -- (-1.25,1);
\draw[thick] (-2.25,-1) -- (-1.75,-0.5);
\draw[thick] (-1.75,-0.5) -- (-1.25,-1);
\draw[thick] (-1.9,0.35) to[out=170, in=-170] (-1.9,-0.35);
\draw[thick] (-1.6,0.35) to[out=10, in=-10] (-1.6,-0.35);
\draw[thick, fill=myblue, rounded corners=2pt] (-2,0.25) rectangle (-1.5,0.75);
\draw[thick, fill=myred, rounded corners=2pt] (-2,-0.25) rectangle (-1.5,-0.75);
\Text[x=-1.25,y=-0.5-\eps]{$j$}
\Text[x=-2.25,y=-0.5-\eps]{$i$}
\Text[x=-2.25,y=+0.5+\eps]{$k$}
\Text[x=-1.25,y=+0.5+\eps]{$\ell$}
\Text[x=-1.25,y=-0.5-\ep]{}
\end{tikzpicture}
=\sum_{p,q}\begin{tikzpicture}[baseline=(current  bounding  box.center), scale=0.6]
\def\ep{1.3};
\def\eps{0.8};
\draw[thick] (-2.25,1) -- (-1.75,0.5);
\draw[thick] (-1.75,0.5) -- (-1.25,1);
\draw[thick] (-2.25,-1) -- (-1.75,-0.5);
\draw[thick] (-1.75,-0.5) -- (-1.25,-1);
\draw[thick] (-1.9,0.35) to[out=170, in=-170] (-1.9,-0.35);
\draw[thick] (-1.6,0.35) to[out=10, in=-10] (-1.6,-0.35);
\draw[thick, fill=myblue, rounded corners=2pt] (-2,0.25) rectangle (-1.5,0.75);
\draw[thick, fill=myred, rounded corners=2pt] (-2,-0.25) rectangle (-1.5,-0.75);
\Text[x=-1.25,y=-0.5-\eps]{$j$}
\Text[x=-2.25,y=-0.5-\eps]{$i$}
\Text[x=-2.25,y=+0.5+\eps]{$k$}
\Text[x=-1.25,y=+0.5+\eps]{$\ell$}
\Text[x=-2.4,y=0]{$p$}
\Text[x=-1.1,y=0]{$q$}
\Text[x=-1.25,y=-0.5-\ep]{}
\Text[x=-0.4,y=0]{$=$}
\draw[thick] (.7,-.75) -- (.7,.75) (1.3,-0.75) -- (1.3,0.75) (.7,-.75) -- (.45,.-1) (1.3,-.75) -- (1.55,.-1) (.7,.75) -- (.45,1) (1.3,.75) -- (1.55,1);
\Text[x=0.5,y=-0.5-\eps]{$i$}
\Text[x=1.55,y=-0.5-\eps]{$j$}
\Text[x=0.5,y=+0.5+\eps]{$k$}
\Text[x=1.55,y=+0.5+\eps]{$\ell$}
\end{tikzpicture},\\
\sum_{p,q} U_{pq}^{k\ell} (U^\dag)_{ij}^{pq} &=& (U U^\dag)_{ij}^{k\ell} = \delta_{ik} \delta_{j\ell} \qquad\Rightarrow\qquad \begin{tikzpicture}[baseline=(current  bounding  box.center), scale=0.6]
\def\ep{1.3};
\def\eps{0.8};
\draw[thick] (-2.25,1) -- (-1.75,0.5);
\draw[thick] (-1.75,0.5) -- (-1.25,1);
\draw[thick] (-2.25,-1) -- (-1.75,-0.5);
\draw[thick] (-1.75,-0.5) -- (-1.25,-1);
\draw[thick] (-1.9,0.35) to[out=170, in=-170] (-1.9,-0.35);
\draw[thick] (-1.6,0.35) to[out=10, in=-10] (-1.6,-0.35);
\draw[thick, fill=myred, rounded corners=2pt] (-2,0.25) rectangle (-1.5,0.75);
\draw[thick, fill=myblue, rounded corners=2pt] (-2,-0.25) rectangle (-1.5,-0.75);
\Text[x=-1.25,y=-0.5-\eps]{$j$}
\Text[x=-2.25,y=-0.5-\eps]{$i$}
\Text[x=-2.25,y=+0.5+\eps]{$k$}
\Text[x=-1.25,y=+0.5+\eps]{$\ell$}
\Text[x=-1.25,y=-0.5-\ep]{}
\end{tikzpicture}=\sum_{p,q}\begin{tikzpicture}[baseline=(current  bounding  box.center), scale=0.6]
\def\ep{1.3};
\def\eps{0.8};
\draw[thick] (-2.25,1) -- (-1.75,0.5);
\draw[thick] (-1.75,0.5) -- (-1.25,1);
\draw[thick] (-2.25,-1) -- (-1.75,-0.5);
\draw[thick] (-1.75,-0.5) -- (-1.25,-1);
\draw[thick] (-1.9,0.35) to[out=170, in=-170] (-1.9,-0.35);
\draw[thick] (-1.6,0.35) to[out=10, in=-10] (-1.6,-0.35);
\draw[thick, fill=myred, rounded corners=2pt] (-2,0.25) rectangle (-1.5,0.75);
\draw[thick, fill=myblue, rounded corners=2pt] (-2,-0.25) rectangle (-1.5,-0.75);
\Text[x=-1.25,y=-0.5-\eps]{$j$}
\Text[x=-2.25,y=-0.5-\eps]{$i$}
\Text[x=-2.25,y=+0.5+\eps]{$k$}
\Text[x=-1.25,y=+0.5+\eps]{$\ell$}
\Text[x=-2.4,y=0]{$p$}
\Text[x=-1.1,y=0]{$q$}
\Text[x=-1.25,y=-0.5-\ep]{}
\Text[x=-0.4,y=0]{$=$}
\draw[thick] (.7,-.75) -- (.7,.75) (1.3,-0.75) -- (1.3,0.75) (.7,-.75) -- (.45,.-1) (1.3,-.75) -- (1.55,.-1) (.7,.75) -- (.45,1) (1.3,.75) -- (1.55,1);
\Text[x=0.5,y=-0.5-\eps]{$i$}
\Text[x=1.55,y=-0.5-\eps]{$j$}
\Text[x=0.5,y=+0.5+\eps]{$k$}
\Text[x=1.55,y=+0.5+\eps]{$\ell$}
\end{tikzpicture},
\end{eqnarray}
while the dual fusion rules amount to two different equivalent expressions of the unitarity of $\tilde{U}$:
\begin{align}
\sum_{pq} (U^\dag)_{kp}^{\ell q} U^{jp}_{iq} =& \sum_{pq} \overline{U^{kp}_{\ell q}} U^{jp}_{iq} = \sum_{pq} \overline{\tilde{U}^{qp}_{\ell k}} \tilde{U}^{qp}_{ij} = \sum_{pq} (\tilde{U}^\dag)^{\ell k}_{qp} {\tilde U}^{qp}_{ij} = (\tilde{U}^\dag \tilde{U})_{ij}^{\ell k} = \delta_{i\ell}\delta_{jk}\notag\\
&\qquad\Rightarrow\qquad\begin{tikzpicture}[baseline=(current  bounding  box.center), scale=0.6]
\def\eps{0.4};
\def\ep{1.6}
\Text[x=-0.125,y=-\ep]{}
\draw[thick] (-2.25+2,1+\eps) -- (-1.75+2,0.5+\eps);
\draw[thick] (-1.75+2,0.5+\eps) -- (-2.25+2,0+\eps);
\draw[thick] (-1.75+2,-0.5-\eps) -- (-2.25+2,0-\eps);
\draw[thick] (-2.25+2,-1-\eps) -- (-1.75+2,-0.5-\eps);
\draw[thick] (-1.6+2,0.7+\eps) to[out=10, in=-10] (-1.6+2,-0.7-\eps);
\draw[thick] (-1.6+2,0.3+\eps) to[out=10, in=-10] (-1.6+2,-0.3-\eps);
\draw[thick, fill=myblue, rounded corners=2pt] (-2+2,0.25+\eps) rectangle (-1.5+2,0.75+\eps);
\draw[thick, fill=myred, rounded corners=2pt] (-2+2,-0.25-\eps) rectangle (-1.5+2,-0.75-\eps);
\Text[x=-.5,y=-1.-\eps]{$i$}
\Text[x=-.5,y=-0.-\eps]{$j$}
\Text[x=-.5,y=+1.+\eps]{$\ell$}
\Text[x=-.5,y=+0.+\eps]{$k$}
\end{tikzpicture}=\sum_{p,q}\begin{tikzpicture}[baseline=(current  bounding  box.center), scale=0.6]
\def\eps{0.4};
\def\ep{1.8}
\Text[x=-0.125,y=-\ep]{}
\draw[thick] (-2.25+2,1+\eps) -- (-1.75+2,0.5+\eps);
\draw[thick] (-1.75+2,0.5+\eps) -- (-2.25+2,0+\eps);
\draw[thick] (-1.75+2,-0.5-\eps) -- (-2.25+2,0-\eps);
\draw[thick] (-2.25+2,-1-\eps) -- (-1.75+2,-0.5-\eps);
\draw[thick] (-1.6+2,0.7+\eps) to[out=10, in=-10] (-1.6+2,-0.7-\eps);
\draw[thick] (-1.6+2,0.3+\eps) to[out=10, in=-10] (-1.6+2,-0.3-\eps);
\draw[thick, fill=myblue, rounded corners=2pt] (-2+2,0.25+\eps) rectangle (-1.5+2,0.75+\eps);
\draw[thick, fill=myred, rounded corners=2pt] (-2+2,-0.25-\eps) rectangle (-1.5+2,-0.75-\eps);
\draw[thick] (-.7+4,0.7+\eps) to[out=10, in=-10] (-.7+4,-0.7-\eps);
\draw[thick] (-.7+4,0.3+\eps) to[out=10, in=-10] (-.7+4,-0.3-\eps);
\draw[thick] (-.7+4,0.7+\eps) -- (-1.05+4,0.7+\eps)(-1.35+4,1+\eps) -- (-1.05+4,0.7+\eps);
\draw[thick] (-.7+4,0.3+\eps) -- (-1.05+4,0.3+\eps)(-1.35+4,+\eps) -- (-1.05+4,0.3+\eps);
\draw[thick] (-.7+4,-0.3-\eps) -- (-1.05+4,-0.3-\eps)(-1.35+4,-\eps) -- (-1.05+4,-0.3-\eps);
\draw[thick] (-.7+4,-0.7-\eps) -- (-1.05+4,-0.7-\eps)(-1.35+4,-1-\eps) -- (-1.05+4,-0.7-\eps);
\Text[x=1.8,y=0]{$=$}
\Text[x=-.5,y=-1.-\eps]{$i$}
\Text[x=-.5,y=-0.-\eps]{$j$}
\Text[x=-.5,y=+1.+\eps]{$\ell$}
\Text[x=-.5,y=+0.+\eps]{$k$}
\Text[x=2.4,y=-1.-\eps]{$i$}
\Text[x=2.4,y=-0.-\eps]{$j$}
\Text[x=2.4,y=+1.+\eps]{$\ell$}
\Text[x=2.4,y=+0.+\eps]{$k$}
\Text[x=1.3,y=0]{$p$}
\Text[x=.55,y=0]{$q$}
\end{tikzpicture}\,,\\
\sum_{p,q} (U^\dag)_{qk}^{p\ell} U_{pi}^{qj} =& \sum_{pq} \overline{U^{qk}_{p\ell}} U_{pi}^{qj} = \sum_{p,q} \overline{\tilde{U}^{\ell k}_{p q}} \tilde{U}_{pq}^{ij} = \sum_{pq}  \tilde{U}_{pq}^{ij}  (\tilde{U}^\dag)_{\ell k}^{pq}= (\tilde{U} \tilde{U}^\dag)_{\ell k}^{ij} = \delta_{i\ell} \delta_{jk}\notag\\
& \qquad\Rightarrow\qquad\begin{tikzpicture}[baseline=(current  bounding  box.center), scale=0.6]
\def\eps{0.4};
\def\ep{1.6}
\Text[x=-0.125,y=-\ep]{}
\draw[thick] (-1.75+2,0.5+\eps) -- (-1.25+2,1+\eps);
\draw[thick] (-1.25+2,0+\eps) -- (-1.75+2,0.5+\eps);
\draw[thick] (-1.25+2,0-\eps) -- (-1.75+2,-0.5-\eps);
\draw[thick] (-1.75+2,-0.5-\eps) -- (-1.25+2,-1-\eps);
\draw[thick] (-1.9+2,0.7+\eps) to[out=170, in=-170] (-1.9+2,-0.7-\eps);
\draw[thick] (-1.9+2,0.3+\eps) to[out=170, in=-170] (-1.9+2,-0.3-\eps);
\draw[thick, fill=myblue, rounded corners=2pt] (-2+2,0.25+\eps) rectangle (-1.5+2,0.75+\eps);
\draw[thick, fill=myred, rounded corners=2pt] (-2+2,-0.25-\eps) rectangle (-1.5+2,-0.75-\eps);
\Text[x=.95,y=-1.-\eps]{$i$}
\Text[x=.95,y=-0.-\eps]{$j$}
\Text[x=.95,y=+1.+\eps]{$\ell$}
\Text[x=.95,y=+0.+\eps]{$k$}
\end{tikzpicture}=\sum_{p,q}\begin{tikzpicture}[baseline=(current  bounding  box.center), scale=0.6]
\def\eps{0.4};
\def\ep{1.8}
\Text[x=-0.125,y=-\ep]{}
\draw[thick] (-1.75+2,0.5+\eps) -- (-1.25+2,1+\eps);
\draw[thick] (-1.25+2,0+\eps) -- (-1.75+2,0.5+\eps);
\draw[thick] (-1.25+2,0-\eps) -- (-1.75+2,-0.5-\eps);
\draw[thick] (-1.75+2,-0.5-\eps) -- (-1.25+2,-1-\eps);
\draw[thick] (-1.9+2,0.7+\eps) to[out=170, in=-170] (-1.9+2,-0.7-\eps);
\draw[thick] (-1.9+2,0.3+\eps) to[out=170, in=-170] (-1.9+2,-0.3-\eps);
\draw[thick, fill=myblue, rounded corners=2pt] (-2+2,0.25+\eps) rectangle (-1.5+2,0.75+\eps);
\draw[thick, fill=myred, rounded corners=2pt] (-2+2,-0.25-\eps) rectangle (-1.5+2,-0.75-\eps);
\draw[thick] (-1.45+4,0.7+\eps) to[out=170, in=-170] (-1.45+4,-0.7-\eps);
\draw[thick] (-1.45+4,0.3+\eps) to[out=170, in=-170] (-1.45+4,-0.3-\eps);
\draw[thick] (-1.1+4,0.7+\eps) -- (-1.45+4,0.7+\eps)(-1.1+4,0.7+\eps) -- (-.75+4,1+\eps);
\draw[thick] (-1.1+4,0.3+\eps) -- (-1.45+4,0.3+\eps)(-1.1+4,0.3+\eps) -- (-.75+4,+\eps);
\draw[thick] (-1.1+4,-0.3-\eps) -- (-1.45+4,-0.3-\eps)(-1.1+4,-0.3-\eps) -- (-.75+4,-\eps);
\draw[thick] (-1.1+4,-0.7-\eps) -- (-1.45+4,-0.7-\eps)(-1.1+4,-0.7-\eps) -- (-.75+4,-1-\eps);
\Text[x=1.45,y=0]{$=$}
\Text[x=.95,y=-1.-\eps]{$i$}
\Text[x=.95,y=-0.-\eps]{$j$}
\Text[x=.95,y=+1.+\eps]{$\ell$}
\Text[x=.95,y=+0.+\eps]{$k$}
\Text[x=3.5,y=-1.-\eps]{$i$}
\Text[x=3.5,y=-0.-\eps]{$j$}
\Text[x=3.5,y=+1.+\eps]{$\ell$}
\Text[x=3.5,y=+0.+\eps]{$k$}
\Text[x=-.8,y=0]{$p$}
\Text[x=-.05,y=0]{$q$}
\end{tikzpicture}\,.
\end{align}
Here $\overline{(\cdots)}$ denotes the complex conjugation.

\section{Properties of $\mathcal M_{\pm}(a)$}
\label{sec:mapsdetails}

In this appendix we prove (i) the spectrum of unistochastic operator map $\mathcal M_{\pm}$ lies on the unit disk ${\mathcal D= \{ \lambda \in \mathbb C; |\lambda| \le 1\}}$; (ii) if an eigenvalue $\lambda$ of $\mathcal M_{\pm}$ lies on the unit circle, i.e., it is unimodular $\lambda=e^{i\theta}$, its algebraic and geometric multiplicities coincide. We start by proving the following identity 
\be
\|\mathcal M_{\pm}(a)\|_1\leq \|a\|_1\,,
\label{eq:contractivenature}
\ee
where $\|A\|_1={\rm tr}[\sqrt{AA^\dag}]$ is the trace norm. Considering for instance $\mathcal M_{+}(a)$, this can be seen as follows  
\be
\|\mathcal M_{+}(a)\|_1= \frac{1}{d} \|{\rm tr}_{1}[U^\dag (a\otimes\1) U]\|_1\leq \frac{1}{d} \|U^\dag (a\otimes\1) U \|_1 = \frac{1}{d} \|a\otimes\1\|_1 = \|a\|_1\,,
\ee
where we used the well known identity (see, e.g., Ref.~\cite{Kitaevbook})
\be
\|{\rm tr}_{1}[A]\|_1 \leq  \|A\|_1\,, \qquad\qquad \forall\, A\in{\rm End}(\mathcal H_1\otimes \mathcal H_2)\,.
\label{eq:identity}
\ee
Considering an eigenvector $a$ of $\mathcal M_{+}(a)$ associated to the eigenvalue $\lambda$, and using  \eqref{eq:contractivenature}, we have 
\be
\|a\|_1\geq \|\mathcal M_{+}(a)\|_1=\| \lambda a\|_1=|\lambda| \|a\|_1 \qquad\Rightarrow\qquad 1 \geq |\lambda|\,.
\ee
This proves the point (i). To prove point (ii) we proceed by \emph{reductio ad absurdum}. Let us assume that (ii) does not hold, then there must exist a generalised eigenvector $b$ of $\mathcal M_{\pm}$ such that 
\be
\mathcal M_{\pm}(b)= e^{i \theta} b+ \alpha a,\qquad \theta\in \mathbb R\,,\qquad\alpha\neq0\,,
\ee 
where $a$ is a proper eigenvector of $\mathcal M_{\pm}$ corresponding to the eigenvalue $e^{i \theta}$. In this case, however, we obtain   
\be
\|\mathcal M_{\pm}^k(b)\|_1 = \|e^{i \theta} b+ k \alpha a\|_1\,, \qquad \forall k\in\mathbb N\,. 
\ee
For large enough $k$ this contradicts \eqref{eq:contractivenature}.

\section{Dual-unitary $4\times 4$ matrices}
\label{app:dualunitarymatrix}

In this section we show that every $4\times 4$ dual-unitary matrices can be written as in Eq.~(23) of the main text. The starting point is to observe that every matrix $O\in{\rm U}(4)$ can be written as (see, e.g., Refs.~\cite{KC:optimalentanglement, VW:optimalquantumcircuits}) 
\be
O=e^{i \phi'} (u'_+\otimes u'_-) V[J_1,J_2,J_3] (v'_-\otimes v'_+)\,,
\label{eq:O}
\ee
where $\phi',J_1,J_2,J_3 \in \mathbb R$, $u_\pm',v'_\pm\in {\rm SU}(2)$ and we defined 
\be
V[J_1,J_2,J_3]=\exp[-i (J_1 \sigma^x\otimes\sigma^x + J_2\sigma^y\otimes\sigma^y+ J_3 \sigma^z\otimes\sigma^z)]\,.
\ee
Considering a matrix $O$ of the form \eqref{eq:O} and computing $\tilde O$ we find 
\be
\tilde O = e^{i \phi'} ((v'_+)^T\otimes u'_-) \tilde V[J_1,J_2,J_3] (v'_-\otimes (u'_+)^T),
\label{eq:dualO}
\ee
where $A^{T}$ denotes the transpose of $A$ and we defined  
\be
 \tilde V[J_1,J_2,J_3]=\begin{bmatrix}
 e^{-i J_3}\cos(J_-) & 0 & 0 & e^{i J_3}\cos(J_+) \\
 0 & -i e^{-i J_3}\sin(J_-)   &   -i e^{i J_3}\sin(J_+)  & 0\\
 0 &  -i e^{i J_3}\sin(J_+)   &  -i e^{-i J_3}\sin(J_-)   & 0\\
 e^{i J_3}\cos(J_+) & 0 & 0 & e^{-i J_3} \cos(J_-) \\
 \end{bmatrix},\qquad\qquad J_\pm=J_1\pm J_2\,. 
\ee
Since transposition preserves unitarity, \eqref{eq:dualO} is unitary if and only if $\tilde V[J_1,J_2,J_3]$ is unitary. The latter condition is realised when  
\be
J_{\sigma(1)}=\frac{\pi}{4}+\frac{\pi}{2}n_1,\qquad J_{\sigma(2)}=\frac{\pi}{4}+\frac{\pi}{2} n_2, \qquad J_{\sigma(3)}=J\,,\qquad n_1,n_2=0,1,2,3,\quad J\in \mathbb R,\quad\sigma\in S_3\,,
\ee
where $S_3$ denotes the group of permutations of three objects. Using ${\rm SU}(2)$ rotations, the permutation $\sigma$ can always be undone. Namely, every dual-unitary matrix $U$ can be written as
\be
U=e^{i \phi'} (u'_+\otimes u'_-) V\left[\frac{\pi}{4}+\frac{\pi}{2}n_1,\frac{\pi}{4}+\frac{\pi}{2}n_2,J\right] (v'_-\otimes v'_+)\,,
\ee
with $u_\pm',v'_\pm\in {\rm SU}(2)$. We then observe
\be
[\sigma^\alpha\otimes\sigma^\alpha,\sigma^\beta\otimes\sigma^\beta]=0,\qquad\qquad\alpha,\beta=x,y,z\,,
\ee
so that 
\be
U=e^{i \phi'} (u'_+\otimes u'_-)  V\left[\frac{\pi}{4},\frac{\pi}{4},J\right] \exp[-i \frac{\pi}{2} n_1 \sigma^x\otimes\sigma^x]\exp[-i \frac{\pi}{2} n_2 \sigma^y\otimes\sigma^y](v'_-\otimes v'_+)\,.
\ee
To conclude we note
\be
\exp[-i \frac{\pi}{2} n \sigma^\alpha\otimes\sigma^\alpha]=\cos\left(\frac{\pi}{2} n\right) \1\otimes\1-i \sin\left(\frac{\pi}{2} n\right) \sigma^\alpha\otimes\sigma^\alpha=
\begin{cases}
(-)^{n/2}\1\otimes\1 & n\quad\text{even}\\
i(-)^{(n+1)/2}\sigma^\alpha\otimes\sigma^\alpha & n\quad\text{odd}
\end{cases}\,.
\label{eq:exppi/2}
\ee 
This means that \eqref{eq:exppi/2} can be written as a product of $SU(2)$ matrices times a phase. We then conclude 
\be
U=e^{i \phi} (u_+\otimes u_-) V\left[\frac{\pi}{4},\frac{\pi}{4},J\right] (v_-\otimes v_+)\,. 
\ee

\section{Quantum circuit formulation of the kicked Ising model}
\label{app:kickedising}

Let us consider the kicked Ising model~\cite{KI_JPA,KI_Ruelle,KI_PRE}, whose dynamics are determined by the following time-periodic Hamiltonian
\be
H_{\rm KI}[t]=H_{\rm I}+\sum_{m=-\infty}^\infty\!\!\!\delta(t-m) H_{\rm K}\,,
\label{eq:ham}
\ee
where we set to one the period of the driving, we denoted the Dirac delta function by $\delta(t)$, and we defined 
\begin{align}
&H_{\rm I}\equiv\! J \sum_{j=0}^{2L-1} \sigma^{z}_j \sigma^z_{j+1}+ h \sum_{j=0}^{2L-1} \sigma^z_{j}\,,\label{eq:hamiltonians1}\\
&H_{\rm K}\equiv b \sum_{j=0}^{2L-1}  \sigma^x_j.
\label{eq:hamiltonians2}
\end{align}
In these equations $2L$ is the volume of the system, the matrices $\sigma_j^a$, with ${a\in\{x,y,z\}}$, are the Pauli matrices at position $j$, and we adopted periodic boundary conditions $\sigma_{2L}^{a}\equiv\sigma_0^{a}$. 

In this appendix we show that the kicked Ising model can be represented as a local quantum circuit. To see this, we write the Floquet operator associated to \eqref{eq:ham}
\be
\mathcal U_{\rm KI}= 
{\frak T}\!\exp\!\!\left[-i\int_0^1\!\!\!\!{\rm d}t\, H_{\rm KI}[t]\right]\!\!=
e^{-i H_{\rm K}}e^{-i H_{\rm I}}\,,
\label{eq:floquet}
\ee
where ${\frak T}\!\exp\left[\cdot\right]$ denotes a time-ordered exponential. Defining 
\begin{align}
U_{\rm I} &= e^{- i J \sigma^{z}\otimes\sigma^z} (e^{- i h \sigma^{z}}\otimes\1), & &\mathbb U^{\rm e}_{\rm I}= U_{\rm I}^{\otimes L}, & &\mathbb U^{\rm o}_{\rm I}= \mathbb T^{\phantom{\dag}}_{2L} \mathbb U^{\rm e}_{\rm I} \mathbb T^\dag_{2L},\\
U_{\rm K} &= e^{- i b \sigma^{x}}, & &\mathbb U_{\rm K}= U_{\rm K}^{\otimes 2L},
\end{align}
and using 
\be
\left[\mathbb U^{\rm o}_{\rm I}, \mathbb U^{\rm e}_{\rm I}\right]=0,
\ee
it is immediate to verify that the integer powers of the Floquet operator can be represented as
\begin{align}
&\mathcal U_{\rm KI}^{2t}= \mathbb U^{\phantom{\rm e}}_{\rm K}\mathbb U^{\rm e}_{\rm I}  \mathbb U_{\rm KI}^t \mathbb U^{\rm o}_{\rm KI} \mathbb U^{\rm e}_{\rm I},\\
&\mathcal U_{\rm KI}^{2t+1}= \mathbb U_{\rm K}\mathbb U^{\rm o}_{\rm I} \mathbb U_{\rm KI}^t \mathbb U^{\rm e}_{\rm I},\qquad\qquad\qquad\qquad t\in\mathbb N\,.
\end{align}
Here $\mathbb U_{\rm KI}$ is the transfer matrix of the unitary circuit with local gate 
\be
U_{\rm KI}=  e^{-i J \sigma^{z}\otimes\sigma^z} (e^{-i h \sigma^{z}} e^{-i b \sigma^{x}}e^{-i h \sigma^{z}}\otimes e^{-i b \sigma^{x}})e^{-i J \sigma^{z}\otimes\sigma^z}\,,
\ee
namely
\begin{align}
\mathbb U_{\rm KI}=\mathbb U^{\rm o}_{\rm KI}\mathbb U^{\rm e}_{\rm KI}, & &\mathbb U^{\rm e}_{\rm KI}=U_{\rm KI}^{\otimes L}=\mathbb U^{\rm e}_{\rm I} \mathbb U_{\rm K}\mathbb U^{\rm e}_{\rm I} , & &\mathbb U^{\rm o}_{\rm KI}= \mathbb T^{\phantom{\dag}}_{2L} \mathbb U^{\rm e}_{\rm KI} \mathbb T^\dag_{2L}.
\end{align}
In particular, at the self dual points 
\be
|J|=|b|= \frac{\pi}{4}\,,
\ee
we have 
\begin{align}
U_{\rm SDKI}&=  (e^{-i h \sigma^{z}} e^{-i b \sigma^{x}}\otimes e^{-i b \sigma^{x}})\cdot e^{-i J \sigma^{y}\otimes\sigma^y} e^{-i J \sigma^{z}\otimes\sigma^z}\cdot (e^{-i h \sigma^{z}}\otimes\1)\\
&=(\sigma^z \otimes\sigma^z)^{\theta_H(-J)}  (e^{-i h \sigma^{z}} e^{i b \sigma^{x}}e^{-i b \sigma^{y}}\otimes e^{i b \sigma^{x}}e^{-i b \sigma^{y}})\cdot e^{-i \frac{\pi}{4}\sigma^{x}\otimes\sigma^x} e^{-i \frac{\pi}{4} \sigma^{y}\otimes\sigma^y}\cdot(e^{i b \sigma^{y}}e^{-i h \sigma^{z}}\otimes e^{i b \sigma^{y}})\,,
\end{align}
where $\theta_H(x)$ is the step function. 

\section{Unimodular eigenvalues of $\mathcal M_\pm$ for $d=2$}
\label{app:polynomial}

In this section we analyse the spectrum of $\mathcal M_\pm$ for $d=2$, identifying all possible occurrences of unimodular eigenvalues. After a similarity transformation both matrices can be brought to the form 
\be
\mathcal M= 
\begin{bmatrix}
1 & 0 & 0& 0 \\
0 & & &\\
0 & & R[w] &\\
0 & & &
\end{bmatrix}\cdot
\begin{bmatrix}
1 & 0 \\
0 & \sin2J & 0 & 0 \\
0&0 &\sin2J & 0\\
0&0 &0 & 1
\end{bmatrix}\,\qquad J\in\left(-\frac{\pi}{4},\frac{\pi}{4}\right],
\label{eq:genericform}
\ee
where $w\in{\rm SU}(2)$ and $R[w]$ denotes is $3\times 3$ adjoin representation (explicitly $R[w]_{\alpha,\beta} = \frac{1}{2}{\rm tr}[\sigma^\alpha w \sigma^\beta w^{-1}]$).
It is then sufficient to analyse the spectrum of \eqref{eq:genericform}. Parametrising $w$ as 
\begin{align}
w=\begin{pmatrix}
r e^{i \phi/2}                     & - \sqrt{1-r^2} e^{-i \theta/2} \\
\sqrt{1-r^2} e^{i \theta/2}         & r e^{- i \phi/2}  
\end{pmatrix},
\qquad\qquad r\in[0,1],\qquad\theta,\phi\in[0,4\pi]\,,
\end{align}
the characteristic polynomial of \eqref{eq:genericform} reads as  
\be
g(\lambda)={\rm det}(\lambda \1 - {\cal M}) = (\lambda-1)p(\lambda)\,,
\ee
where 
\be
p(\lambda)=\left[\lambda ^3 + \lambda ^2 \left( 1-2 r^2 (\sin 2 J \cos\phi+1) \right) +  \lambda  \sin 2 J \left(2 r^2 (\sin 2 J+ \cos\phi)- \sin 2 J\right)  - \sin ^2 2 J\right].
\ee
Apart from the trivial root $\lambda=1$ coming from the decoupled identity sector, other unimodular roots of $g(\lambda)$ can appear at some special values of the parameters. Note that $g(\lambda)$ does not depend on $\theta$. This, as mentioned in the main text, is a direct consequence of the invariance of $\mathcal M[J]$ under rotations around the $z$ axis. Moreover, we also see that $g(\lambda)$ is $2\pi$-periodic in $\phi$, so that we can restrict our analysis to $\phi\in[0,2\pi]$

Since $p(\lambda)$ is a polynomial with real coefficients, it has either three real roots or one real root and a complex conjugated pair. This means that we are interested in the cases where we have: (i) real solutions equal to $1$; (ii) real solutions equal to $-1$; (iii) complex conjugated  pairs of solutions with unit magnitude. 

Let us start considering the case (i), imposing $p(1)=0$ we have 
\begin{align}
 (r^2 - 1)(\sin^22J - 1) = 0,
\end{align}
which is solved for $r=1$ or $J={\pi}/{4}$. In particular, for $r=1$ the polynomial reads as 
\be
p(\lambda)=(\lambda-1)(\lambda^2-2\lambda \cos\phi\sin2J+\sin^22J)
\ee
so that the two other solutions are $\lambda_{\pm}=e^{\pm i \phi} \sin2J$. Similarly, for the case (ii) we impose $p(-1)=0$ and obtain the condition 
\begin{align}
r (1 + 2 \cos\phi \sin 2 J + \sin^22 J) = 0,
\end{align}
which is solved for $r=0$, or ($J={\pi}/{4}$, $\phi=\pi$). In particular, for $r=0$ the polynomial reads as 
\be
p(\lambda)=(\lambda+1)(\lambda^2-\sin2J)
\ee
so that the two other solutions are $\lambda_\pm=\pm \sin2J$. Finally, considering the case (iii) and demanding the presence of a pair of conjugated solutions $e^{ \pm i h}$ we have  
\begin{align}
& 4 r^2 \sin (2 J) \cos ( \phi )+\cos (4 J)+4 r^2-3-4 \cos h =0, \\
& r^2 (4 (1-2 \cos h ) \sin (2 J) \cos ( \phi )-2 \cos (4 J)+2-8 \cos h )+\cos (4 J)-3+8 \cos h ^2+4 \cos h = 0.
\end{align} 
These conditions are obtained by requiring that $(\lambda-e^{i h})(\lambda - e^{- i h})$ divides the characteristic polynomial. Expressing $ \cos h$ from the first equation and inserting it into the second leads to
\begin{align}
\cos ^2(2 J) \left(4 r^2 \sin (2 J) \cos ( \phi )+\cos (4 J)-3\right) =0,
\end{align}
which is solved only for $J = {\pi}/{4}$.
In this case, the phase of the complex solutions reads as
 \begin{align}
 \cos h = r^2 (\cos\phi + 1)-1
 \end{align}
 and the polynomial is written as 
 \be
 p(\lambda)=(\lambda-1)(\lambda-e^{i h})(\lambda+e^{i h})\,.
 \ee
 Namely, the third root is equal to one.

Putting all together, as a function of the parameters $J,r,$ and $\phi$, we found the following cases  
\begin{align}
& J=\pi/4\land\begin{cases}
 \phi = \pi, \quad \forall r: &(\lambda_1=1,\lambda_2=-1,\lambda_3=-1)\\
 r = 0, \quad \forall \phi: &(\lambda_1=1,\lambda_2=-1,\lambda_3=-1)\\
 \phi \neq \pi\land r=1/\sqrt{\cos\phi+1}: &(\lambda_1=1,\lambda_2=1,\lambda_3=1)\\
 \phi \neq \pi \land r\neq0,1/\sqrt{\cos\phi+1}: &(\lambda_1=1,\lambda_2=e^{-i h},\lambda_3=e^{i h})\quad \text{with}\quad h \neq 0,\pi\,.\\
 \end{cases}\\
& J\neq \pi/4\land \begin{cases}
r = 1,\quad \forall \phi: & (\lambda_1=1,  \lambda_2=e^{i \phi} \sin2J,  \lambda_3=e^{-i \phi} \sin2J)\\
r = 0,\quad \forall \phi: & (\lambda_1=-1, \lambda_2=\sin2J,\lambda_3=-\sin2J)\\
r\neq0,1,\quad \forall \phi: & (\lambda_1=a,\lambda_2=b,\lambda_3=c) \quad \text{with}\quad |a|,|b|,|c|<1\,.\\
\end{cases}
 \end{align}
 Noting that both $\mathcal M_+$ and $\mathcal M_-$ depend on the parameter $J$, while they depend on two independent parameters $r_\pm$ and $\phi_\pm$, we can single out the following distinct classes (in the notation introduced in the main text) 
\begin{itemize}
\item $J=\pi/4$ $\Rightarrow$ (i), (ii); 
\item $J\neq\pi/4$ $\Rightarrow$ (ii), (iii), (iv); 
\end{itemize}  


\begin{thebibliography}{99}

\bibitem{altland} A.~Altland~and~B.~Simons, \emph{Condensed Matter Field Theory}, Cambridge University Press (2010).

\bibitem{sethna} J.~P.~Sethna, \emph{Statistical Mechanics: Entropy, Order Parameters, and Complexity}, Oxford University Press (2006).

\bibitem{arnold} V.~I.~Arnold and A.~Avez, \emph{Ergodic Problems of Classical Mechanics}, Addison-Wesley, Reprint Edition (1989).

 \bibitem{mahanbook}
G. D. Mahan, \emph{Condensed Matter in a Nutshell}, Princeton University Press (2011). 

\bibitem{blockreview}
I.~Bloch,~J.~Dalibard,~and~W.~Zwerger, \href{https://doi.org/10.1103/RevModPhys.80.885}{Rev. Mod. Phys. {\bf 80}, 885 (2008)}.

\bibitem{korepin} V.~E.~Korepin, N.~M.~Bogoliubov, and A.~G.~Izergin, \emph{Quantum Inverse Scattering Method and Correlation Functions},  Cambridge Universiity Press (1997).

\bibitem{cornfeld} I.~P.~Cornfeld,  S.~V.~Fomin, Y.~G.~Sinai, \emph{Ergodic Theory}, Springer (2012).

\bibitem{gaspard} P.~Gaspard, \emph{Chaos, Scattering and Statistical Mechanics}, Cambridge University Press (1998).

\bibitem{ott} E.~Ott, \emph{Chaos in Dynamical Systems}, 2nd Edition, Cambridge University Press (2012).

\bibitem{ColinDeVerdiere} Y.~C.~De Verdiere, \href{https://doi.org/10.1007/BF01209296}{Commun. Math. Phys. {\bf 102}, 497 (1985).}

\bibitem{Zelditch} S.~Zelditch, \href{https://projecteuclid.org/euclid.dmj/1077306306}{Duke Math. J. {\bf 55},  919 (1987).}

\bibitem{DeBievre} A.~Bouzouina~and~S.~De Bievre, \href{https://DOI.org/10.1007/BF02104909}{Commun. Math. Phys. {\bf 178}, 83 (1996).}


\bibitem{TP_PRE99} T.~Prosen, \href{https://journals.aps.org/pre/pdf/10.1103/PhysRevE.60.3949}{Phys. Rev. E {\bf 60}, 3949 (1999).}
	 
\bibitem{KI_JPA} T.~Prosen, \href{https://doi.org/10.1088/1751-8113/40/28/S02}{J. Phys. A {\bf 40}, 7881 (2007)}.
	 
\bibitem{ehud} D.~E.~Parker, X.~Cao, T.~Scaffidi and E.~Altman, \href{https://arxiv.org/abs/1812.08657}{\tt arXiv:1812.08657 (2018).}
	 
\bibitem{graffi} S.~Graffi~and~A.~Martinez,~\href{https://doi.org/10.1063/1.531741}{J. Math. Phys. {\bf 37}, 5111 (1996)}.
	 
\bibitem{lenci} M. Lenci, \href{https://doi.org/10.1063/1.531684}{J. Math. Phys. {\bf 37}, 5137 (1996)}.

\bibitem{letter} 
B.~Bertini,~P.~Kos,~and~T.~Prosen, \href{https://doi.org/10.1103/PhysRevLett.121.264101}{Phys. Rev. Lett. {\bf 121}, 264101 (2018)}. 

\bibitem{entropy} 
B.~Bertini,~P.~Kos,~and~T.~Prosen, \href{https://doi.org/10.1103/PhysRevX.9.021033}{Phys. Rev. X {\bf 9}, 021033 (2019)}. 

\bibitem{PoGri17}
V. Gritsev and A. Polkovnikov, \href{https://doi.org/10.21468/SciPostPhys.2.3.021}{SciPost Phys. {\bf 2}, 021 (2017)}.

\bibitem{Lenart1}
M. Vanicat, L. Zadnik, and T. Prosen, \href{https://doi.org/10.1103/PhysRevLett.121.030606 }{Phys. Rev. Lett. 121, 030606 (2018)}.

\bibitem{Lenart2}
M. Ljubotina, L. Zadnik, and T. Prosen, \href{https://doi.org/10.1103/PhysRevLett.122.150605}{Phys. Rev. Lett. {\bf 122}, 150605 (2019)}.

\bibitem{SM} See the Supplemental Material, which includes Refs.~\cite{Kitaevbook, KI_PRE, KI_Ruelle, KC:optimalentanglement, VW:optimalquantumcircuits}, for some useful complementary information.

	

\bibitem{Kitaevbook}
A.~Yu.~Kitaev,~A.~H.~Shen,~M.~N.~Vyalyi, \emph{Classical and quantum computation}, AMS (2002). 

\bibitem{KI_PRE} T.~Prosen, \href{https://doi.org/10.1103/PhysRevE.65.036208}{Phys. Rev. E, {\bf 65}, 036208 (2002)}.

\bibitem{KI_Ruelle} 
T.~Prosen, \href{http://iopscience.iop.org/article/10.1088/0305-4470/35/48/102/meta}{J. Phys. A {\bf 35}, L737 (2002)}.

\bibitem{KC:optimalentanglement} B. Kraus and J. I. Cirac, \href{https://doi.org/10.1103/PhysRevA.63.062309}{Phys. Rev. A {\bf 63}, 062309 (2001)}.

\bibitem{VW:optimalquantumcircuits} F. Vatan and C. Williams, \href{https://doi.org/10.1103/PhysRevA.69.032315}{Phys. Rev. A {\bf 69}, 032315 (2004)}. 
	

\bibitem{footnote} Recently, Ref.~\cite{austen} independently showed that dual-unitarity implies a flat entanglement spectrum.

\bibitem{austen}
S.~Gopalakrishnan~and~A.~Lamacraft, \href{https://doi.org/10.1103/PhysRevB.100.064309}{Phys. Rev. B 100, 064309 (2019)}.	
	
	

\bibitem{ZB:quantummaps} K. \.Zyczkowski and I. Bengtsson, \href{https://doi.org/10.1023/B:OPSY.0000024753.05661.c2}{Open Syst. Inf. Dyn. {\bf 11}, 3-42 (2004)}. 

\bibitem{ZB:book} K. \.Zyczkowski and I. Bengtsson, \emph{Geometry Of Quantum States: An Introduction To Quantum Entanglement}, Cambridge University Press (2017). 

\bibitem{MKZ:unistochastic} M. Musz, M. Ku\'s, and K. \.Zyczkowski, \href{https://doi.org/10.1103/PhysRevA.87.022111}{Phys. Rev. A {\bf 87}, 022111 (2013)}. 



\bibitem{note2}
Note that the parameter $J$ fully specifies the entangling power $\rm EP$ of $U$ (see, e.g., \cite{VW:optimalquantumcircuits}). Specifically we find ${\rm EP}(U)=(2/9)\cos(2J)$. 


\bibitem{RandomCircuitsEnt} 
A. Nahum, J. Ruhman, S. Vijay, and J. Haah, \href{https://doi.org/10.1103/PhysRevX.7.031016}{Phys. Rev. X {\bf 7}, 031016 (2017)}. 
	
\bibitem{Keyserlingk}
C. W. von Keyserlingk, T. Rakovszky, F. Pollmann, and S. L. Sondhi, \href{https://doi.org/10.1103/PhysRevX.8.021013}{Phys. Rev. X {\bf 8}, 021013 (2018)}. 	
	
\bibitem{Chalker} 
A.~Chan,~A.~De Luca,~J.~T.~Chalker,~\href{https://doi.org/10.1103/PhysRevX.8.041019}{Phys. Rev. X {\bf 8}, 041019 (2018)}.

\bibitem{RPK:conservationlaws}
T.~Rakovszky,~F.~Pollmann,~C.~W.~von~Keyserlingk, \href{https://doi.org/10.1103/PhysRevLett.122.250602}{Phys. Rev. Lett. {\bf 122}, 250602 (2019)}.

\bibitem{H:conservationlaws}
Y. Huang,~\href{https://arxiv.org/abs/1902.00977}{\tt arXiv:1902.00977 (2019)}. 
			
					
\end{thebibliography}
\end{document}